\documentclass[aip,jcp,preprint]{revtex4-1}

\usepackage{graphicx}
\usepackage{amsmath,amssymb,amsfonts}
\usepackage{etoolbox}
\usepackage{xcolor}

\newcommand{\bGamma}{\boldsymbol{\Gamma}}
\newcommand{\bC}{\mathbf{C}}
\newcommand{\bv}{\mathbf{v}}

\newcommand{\ave}[1]{\langle #1 \rangle}

\begin{document}

\title{Fractional Viscoelasticity in Transient Unentangled Polymer Networks}
\author{Sachin Shanbhag}
\email{sshanbhag@fsu.edu}
\affiliation{Department of Scientific Computing, Florida State University, Tallahassee, Florida 32306, United States}%

\author{Ralm G. Ricarte}
\email{rricarte@eng.famu.fsu.edu}
\affiliation{Department of Chemical and Biomedical Engineering, Florida A\&M University - Florida State University College of Engineering, Tallahassee, Florida 32310, United States}%

\date{\today}

\begin{abstract}

Stress relaxation in transient polymer networks often shows extended power-law behavior, $G(t) \sim t^{-\beta}$, where the exponent $\beta$ frequently departs from the value $1/2$ predicted by the sticky Rouse model and its variants. We introduce the fractional inhomogeneous Rouse model (FIRM), which uses a generalized Langevin equation driven by fractional Gaussian noise of exponent $\alpha$, while retaining heterogeneous bead friction to represent sticky cross-links. Thus, FIRM unifies subdiffusive sticker dynamics and chain heterogeneity within a single framework. We show that the relaxation modulus $G(t)$ can be represented as a linear combination of Mittag-Leffler functions. For homogeneous chains, it recovers two power-law regimes, $t^{-\alpha/2}$ and $t^{-2\alpha}$, on either side of the terminal relaxation time. Fitting FIRM to stress relaxation data for an imine-based polystyrene vitrimer shows that $\alpha < 1$ is required to capture the shape of the terminal relaxation. It also accommodates both Arrhenius and non-Arrhenius temperature dependence in the rheological activation energy. We derive expressions for dynamic properties such as mean-squared displacement and dielectric response  and outline how generalized memory kernels extend the framework to real materials. Together, these results suggest novel ways in which data from rheology, dielectric spectroscopy, scattering, and other experimental methods may be incorporated into a chemistry-specific molecular model.
\end{abstract}

\maketitle

\section{Introduction}

\subsection{Motivation}

The transient polymer network, in which reversible cross-links form intermolecular connections between chains, is a classical concept in polymer physics. Stern and Tobolsky introduced this idea in 1946 to explain the unusual rheological behavior of polysulfide rubbers. Unlike other synthetic rubbers of that era, which were viscoelastic solids with fixed shapes, polysulfide rubbers behaved as viscoelastic liquids that flowed at elevated temperatures. Stern and Tobolsky attributed this behavior to covalent sulfide bonds that were not permanent but instead reversibly exchanged between chains.\cite{Stern1946}  That same year, Green and Tobolsky formalized this concept by extending rubber elasticity theory to account for continuous cross-link formation and dissociation during network deformation.\cite{Green1946} Over the next 80 years, the polymer physics community expanded this framework to many synthetic and biological transient polymer networks, in which supramolecular, electrostatic, or dynamic covalent bonds form the reversible cross-link junctions.\cite{Tanaka_2011, Webber2022, Ricarte2026}

Vitrimers represent a class of transient polymer networks that has received considerable attention in recent years. These materials consist of polymer chains decorated with dynamic covalent bonds that undergo associative exchange reactions. Under this mechanism, a cross-link forms a new covalent bond with a free reactive site before it breaks from its original junction partner. Associative cross-link exchange enables topology fluctuations while preserving a constant number of cross-links, even as the network flows. As a result, vitrimers combine two properties previously considered to be mutually exclusive: insolubility in good solvents and melt processability at elevated temperatures. This paradoxical marriage between connectivity constraints and structural fluidity imparts vitrimers with the mechanical robustness of thermosets while preserving the reprocessability of thermoplastics, inspiring their development for applications ranging from recyclable composites to shock wave absorbance.\cite{Montarnal2011, Winne2019, McBride2019, VanZee2020, Hayashi2025, Ricarte2026} 

The coupling between network strand relaxation and cross-link exchange gives vitrimers a rich linear viscoelastic response. Their relaxation spectra generally contain fast and slow regimes. The fast regime arises from segmental motion along the polymer backbone and typically exhibits a temperature dependence well described by the Williams-Landel-Ferry (WLF) expression.\cite{Williams1955}  In contrast, the slow regime depends on cross-link exchange and often follows an apparent Arrhenius temperature dependence with a constant activation energy $E_{a}$. However, local cross-link exchange events alone do not appear to fully govern this slow relaxation. Its underlying mechanism remains an open question and has attracted significant attention over the past fifteen years.\cite{Ricarte2026, Meng2022, Lin2025, Wang2026, Quinteros-Sedano2026}

A distinctive feature of the slow relaxation regime, as observed in multiple rheological measurements, is an extended power-law that persists over many decades. Shanbhag, Ricarte, and coworkers reported this behavior in several vitrimer systems, finding that the stress relaxation modulus decays as $G(t) \sim t^{-\beta}$, where $\beta$ depends on the vitrimer chemistry. Dioxaborolane-based polybutadiene vitrimers exhibited $\beta = 0.4$, whereas imine-based polystyrene vitrimers exhibited $\beta = 1.6$. \cite{Ricarte2023, Barzycki2025, Barzycki2026} Nicola\"{y} and coworkers reported similar power-law relaxation in dioxazaborocane-based methacrylic vitrimers, with $\beta \sim 0.4 - 0.7$.\cite{Quinteros-Sedano2025} Sokolov and coworkers observed an even broader scaling regime in a disulfide-based polydimethylsiloxane vitrimer, where the complex modulus followed $G^{*} \sim \omega^{0.2}$ over more than 15 decades of angular frequency $\omega$.\cite{Martins2023} This collection of observations demonstrates that the exponent $\beta$ varies across systems and does not always coincide with the value $1/2$ predicted by variants of the Rouse model.\cite{rouse53, Baxandall1989} Understanding the molecular origin of these power-laws, and in particular their dependence on network architecture and chain dynamics is an important open question.

The theory most commonly used to interpret the linear rheology of unentangled transient networks is the sticky Rouse model and its generalizations.\cite{Baxandall1989, Leibler1991, Rubinstein1998, Chen2013, Chen2016} In this framework, a chain is modeled as a sequence of $N$ beads connected by linear springs in which a subset of beads --- those that participate in temporary cross-links --- are designated as ``sticky''.  When sticky beads are sufficiently numerous and sluggish,\cite{Ricarte2021} the slow and fast relaxation modes are well separated. In this regime, the sticky Rouse model (SRM)  asserts that $G(t)$ can be partitioned into two independent Rouse-like contributions:
\begin{equation}
\frac{G_\text{SRM}(t)}{G_0} = \frac{1}{N}\left[
  \sum_{p=1}^{N_x} \exp\!\left(-\frac{p^2 t}{\tau_x N_x^2}\right)
+ \sum_{p=N_x+1}^{N} \exp\!\left(-\frac{p^2 t}{\tau N^2}\right)
\right],
\label{eqn:SRM}
\end{equation}
where $G_0$ is the modulus and $N_x \gg 1$ is the number of sticky beads per chain. $\tau_x$ and  $\tau$ are timescales that characterize the mobility of the sticky and regular Rouse beads, respectively. The inhomogeneous Rouse model (IHR) relaxes several assumptions in the SRM by numerically evaluating the full spectrum of relaxation times.\cite{Ricarte2021} Although it retains the mathematical structure of Eq.~\ref{eqn:SRM}, it correctly captures interactions that  arise from the interplay between fast (segmental motion) and slow (cross-link exchange) relaxation modes.\cite{Jiang2021, Shao2022, Cui2023, Cui2025, Zhang2026}

A critical feature of Eq.~\ref{eqn:SRM} is that each relaxation mode decays exponentially. Like the original Rouse model,\cite{rouse53, doipd} the superposition of the exponential modes in the SRM and IHR permits a single power-law behavior during transition regimes, i.e., $G(t) \sim t^{-1/2}$. Thus, the SRM and IHR are structurally incapable of explaining experimentally observed power-law exponents that deviate from $\beta = 1/2$. This discrepancy leads to a natural question: what molecular mechanism could give rise to a different power-law exponent in the first place, and how should the sticky Rouse framework be modified to accommodate it?

\subsection{Generalized Langevin Equation}

A particle suspended in a structureless medium undergoes ordinary Brownian motion --- a Markovian, memoryless process --- and its mean-squared displacement (MSD) grows linearly in time, $\rho(t)\sim t$. These dynamics are captured by the Langevin equation in which the particle experiences an instantaneous frictional drag and an uncorrelated stochastic force that satisfy the fluctuation-dissipation theorem.\cite{Langevin1908, doipd}

When the medium surrounding the particle is itself viscoelastic this Markovian description breaks down. A thermal fluctuation that displaces a monomer against its viscoelastic surroundings generates a restoring force that decays over a finite timescale. Successive displacements are therefore correlated, and the resulting dynamics are anomalous: typically, the MSD grows as $\rho(t) \sim t^\alpha$.\cite{Mori1965a, Zwanzig1973, Lee2000, McKinley2009, McKinley2018} Passive microrheology exploits this connection to characterize the linear viscoelasticity of complex fluids by tracking the MSD of particle probes.\cite{Mason1995, Mason1997, Fricks2009, Cordoba2012}

The generalized Langevin equation (GLE) formalizes this picture. In the overdamped limit, it replaces the instantaneous friction of the ordinary Langevin equation with a convolution:
\begin{equation}
\int_0^t K(t - t')\, \dot{\mathbf{r}}(t')\, dt' = \mathbf{F}_\text{ext}(t) + \mathbf{f}(t),
\label{eqn:GLE_intro}
\end{equation}
where $\mathbf{F}_\text{ext}(t)$ is the external force acting on the particle located at $\mathbf{r}$,  $\dot{\mathbf{r}}$ is the time derivative of $\mathbf{r}$, $K(t)$ is a stationary memory kernel that encodes temporal correlations in the friction, and $\mathbf{f}(t)$ is a stochastic force that satisfies the modified fluctuation-dissipation theorem $\langle \mathbf{f}(t) \cdot \mathbf{f}(t') \rangle \propto K(t - t')$. Using projection operator techniques, Mori and Lee showed  that the GLE arises naturally from the exact equations of motion of an interacting many-body system.\cite{Mori1965a, Lee2000} The key insight is that integrating out the fast degrees of freedom of the bath generates a non-Markovian memory in the equation of motion for the slow variables of interest.

A tractable and physically motivated choice for the kernel is $K(t) \sim t^{-\alpha}$ with $0 < \alpha \leq 1$, which corresponds to fractional Gaussian noise (fGn).\cite{Mandelbrot1968} For this kernel, the MSD of a particle governed by Eq.~\ref{eqn:GLE_intro} is subdiffusive and scales as $t^\alpha$.\cite{Barkai2000, Shanbhag2023} Subdiffusive behavior ($\alpha < 1$) is common in polymer systems.\cite{doipd, Weber1993, Wong2004, Tang2015a, Shanbhag2023} Panja showed that the equation of motion of the central monomer of a Rouse chain follows a GLE with $K(t) \sim t^{-1/2}\,e^{-t/\tau_R}$, where $\tau_R$ is the Rouse relaxation time.\cite{Panja2010} Tian et al.\ derived an explicit analytical form for the memory kernel of the end-to-end vector of a Rouse chain.\cite{Tian2022} This expression simplifies to $K(t) \sim t^{-1/2}$ at intermediate times before terminal relaxation ($t \gtrsim \tau_R$). Kou and Xie used the GLE with fGn to explain the equilibrium fluctuation of the distance between an electron donor and acceptor pair within a single protein molecule.\cite{Kou2004} These examples show that the GLE with a power-law memory kernel is not merely a phenomenological ansatz but also has microscopic underpinnings in polymer dynamics.

The GLE perspective also provides a natural starting point for understanding the viscoelasticity of polymer melts. The renormalized Rouse model of Schweizer posits that the dynamics of a tagged chain in a melt of other chains are governed by a GLE analogous to Eq.~\ref{eqn:GLE_intro}, where the memory kernel encodes the viscoelastic response of the surrounding matrix.\cite{Schweizer1989, Schweizer1989a, Schweizer1991, Schweizer1997} Sharma and Cherayil showed that when this kernel takes the fGn form, $G(t) \sim t^{-\beta}$ over intermediate timescales. The stress relaxation exponent $\beta$ is related to the fGn exponent $\alpha$ via $\beta = \alpha/2$.\cite{Sharma2010, sharma2018theoretical} We label this special case as the \textit{fractional Rouse model} (FRM) in which the microscopic origin of fractional viscoelasticity traces back to power-law memory kernel.

\subsection{Goals}

The present work is thus motivated by the convergence of two observations. First, the stress relaxation of unentangled vitrimer melts and other transient polymer networks frequently exhibits power-law relaxation with an exponent that differs from the $t^{-1/2}$ prediction of the sticky Rouse model (SRM or IHR).\cite{Ricarte2021} Second, the FRM provides a principled molecular mechanism that can produce such generalized power-laws but has so far been applied only to homogeneous chains with identical monomers.\cite{Sharma2010}

\begin{figure}
\begin{center}
\includegraphics[scale=0.6]{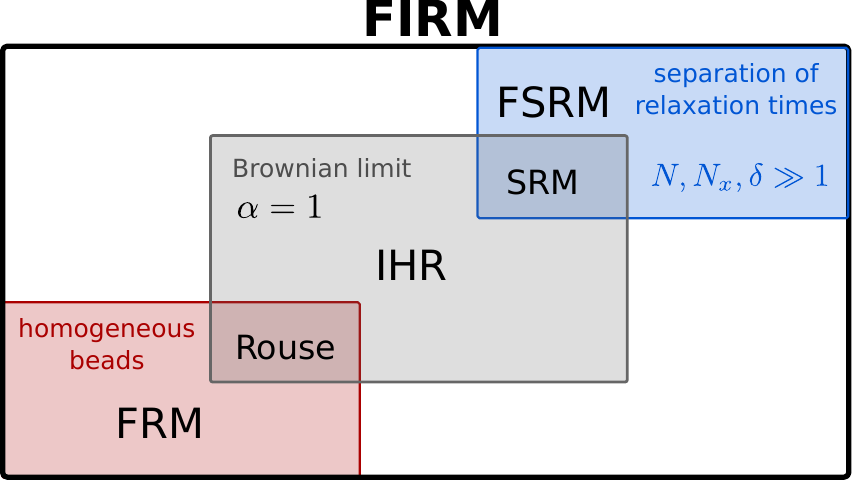}
\end{center}
\caption{Special limits of FIRM: homogeneity (FRM, standard Rouse model), Brownian dynamics (Rouse, IHR, SRM), and separation of slow and fast relaxation timescales (SRM, FSRM). The gray box in the center denotes the $\alpha = 1$ limit, while red box in the lower left corner denotes homogeneous chains.\label{fig:models}}
\end{figure}

Our goal is to marry these two frameworks. We generalize the IHR model by replacing the Langevin equation describing the sticky dynamics with a GLE governed by a fGn memory kernel. The resulting \emph{fractional inhomogeneous Rouse model} (FIRM) allows beads to carry different friction coefficients --- as required to model sticky cross-links --- while simultaneously accounting for the subdiffusive, non-Markovian dynamics that arise when the chain is embedded in a viscoelastic medium. In the special case of a homogeneous chain, FIRM reduces to FRM. In the Markovian limit ($\alpha = 1$) it reduces to the IHR model. FIRM therefore provides a unified framework that smoothly interpolates between these limiting cases and enables the exploration of how subdiffusive matrix dynamics and network cross-linking jointly shape the linear viscoelasticity of unentangled vitrimers (see figure \ref{fig:models}).

\section{Methods}

While the Rouse model is attributed to Rouse,\cite{rouse53} closely related ideas were simultaneously developed by Soviet scientists.\cite{Kargin1949, Gotlib1952, Likhtman2012} The Rouse model neglects both excluded-volume effects and hydrodynamic interactions to yield an analytically tractable description of viscoelastic relaxation and diffusion which corresponds surprisingly well with experiments on unentangled polymer melts.\cite{Ferry1980}

We start by recounting the FRM which generalizes the original Rouse model by assuming that beads are buffeted by fGn instead of Brownian kicks. Subsequently, we account for beads with different drag coefficients and develop the FIRM.

\subsection{Fractional Rouse Model}

Consider the motion of a tagged linear polymer chain with $N$ monomers, embedded in a sea of other chains. Since the environment of the tagged chain is viscoelastic, its evolution can be modeled by an over-damped GLE,\cite{Schweizer1997, Sharma2010}
\begin{equation}
\zeta  \int_0^t dt'\, K(t-t')\, \frac{\partial \mathbf{r}(n,t')}{\partial t'}
  = k \dfrac{\partial^2 \mathbf{r}}{\partial n^2} + \mathbf{f}(t), 
\label{eqn:GLE_Rouse}
\end{equation}
where $\mathbf{r}(n, t)$ denotes the position of the $n$th bead at time $t$, $K$ is a memory kernel that encodes the decay of random fluctuations imposed by the surrounding viscoelastic matrix, and $\zeta$ is a generalized drag or monomer friction coefficient. Beads are connected via linear springs with spring constant $k =3k_BT /b^2$, where $k_B T$ is thermal energy and $b$ is the statistical segment length. We assume that $N$ is sufficiently large to invoke the continuum limit, but sufficiently small to avoid entanglements.

Random kicks on different parts of the chain are assumed to be spatially uncorrelated. The random force implied by such a spatially local kernel obeys a fluctuation-dissipation theorem given by
\begin{equation}
  \langle \mathbf{f}(t) \cdot \mathbf{f}(t')\rangle = 3 \zeta k_BT\, K(t-t').
\label{eqn:fdt_gle_rouse}
\end{equation}
The prefactor of three arises because the polymer chain lives in three-dimensional space. Furthermore, we assume that memory kernel
\begin{equation}
K(t-t') = \dfrac{(t-t')^{-\alpha}}{\Gamma(1-\alpha)}
\label{eqn:fgn_kernel}
\end{equation}
is consistent with fGn. This implies that the units of $\zeta$, kg$\cdot$s$^{\alpha-2}$, depend on $\alpha$ so that the units of the product $\zeta K(\cdot)$, kg$\cdot$s$^{-2}$, are independent of $\alpha$. In this case, the LHS of Eq.~\ref{eqn:GLE_Rouse} can be written in terms of the Caputo fractional derivative as $\zeta \left( \partial^{\alpha} \mathbf{r}/\partial t^{\alpha} \right)$.\cite{Caputo1967}

Note that Eqs.~\ref{eqn:GLE_Rouse} and \ref{eqn:fdt_gle_rouse} reduce to the Langevin equation describing the standard Rouse model when $K(t - t') = 2 \delta(t-t')$. When the delta function is centered on the upper limit of integration, $\int_0^{t} \delta(t - t') f(t)\, dt' = f(t)/2$ for any function $f(\cdot)$. Thus, the LHS in Eq.~\ref{eqn:GLE_Rouse} simplifies to a standard derivative 
\begin{equation}
\zeta  \frac{\partial \mathbf{r}(n,t)}{\partial t}
  = k \dfrac{\partial^2 \mathbf{r}}{\partial n^2} + \mathbf{f}(t). 
\label{eqn:LE_Rouse}
\end{equation}
Interestingly, it can be shown that the fGn kernel with $\alpha = 1$ is equivalent to a memoryless kernel in the sense of distributions acting under a causal time integral. This implies that the FRM recovers the standard Rouse model in the $\alpha = 1$ limit as anticipated by the Caputo derivative.

Eq.~\ref{eqn:GLE_Rouse} is a stochastic integro-differential equation. The dependence on $n$ and $t$ can be decoupled via normal modes $\mathbf{X}_p$ for $p \in [0, N]$ as
\begin{equation}
\mathbf{X}_p(t) = \frac{1}{N} \int_{0}^{N} \mathbf{r}(n, t) \cos \left( \frac{p \pi n}{N} \right) dn
\label{eqn:normal_modes_definition}
\end{equation}
The mode $p=0$ corresponds to the center-of-mass motion, while $p \ge 1$ represents internal relaxation modes. Substituting $\mathbf{X}_p$ into Eq.~\ref{eqn:GLE_Rouse} yields a GLE in the normal modes
\begin{equation}
\int_0^t dt'\, K(t-t')\, \dot{\mathbf{X}}_p(t') + k_p \mathbf{X}_p(t) = \mathbf{F}_p(t), 
\label{eqn:GLE_Xp}
\end{equation}
where $k_p = k p^2 \pi^2/\zeta N^2$. The random force $
\mathbf{F}_p(t) = (\zeta N)^{-1} \int_{0}^{N} \cos (p \pi n/N)\, \mathbf{f}(t)\, dn$ is also consistent with fGn. Thus, $\langle \mathbf{F}_p(t) \rangle = \mathbf{0}$, and the corresponding fluctuation-dissipation theorem is
\begin{equation}
  \langle \mathbf{F}_p(t) \cdot \mathbf{F}_q(t')\rangle = 3 k_BT C_p^2\, \delta_{pq} K(t-t'),
\label{eqn:fdt_gle_normalmodes}
\end{equation}
where $C_p = 1/\sqrt{2 N\zeta}$ for $p \geq 1$ and $C_0 = 1/\sqrt{N\zeta}$. The relaxation modulus,\cite{Sharma2010}
\begin{equation}
\dfrac{G(t)}{G_0} = \dfrac{1}{N}
  \sum_{p=1}^{N} E_\alpha^2\!\left(-(t/\tau_p)^\alpha\right),
  \label{eqn:frm_Gt}
\end{equation}
where $G_0 = ck_BT = G(t=0)$ is proportional to the polymer concentration $c$. $E_\alpha(\cdot)$ is the one-parameter Mittag-Leffler function. For $\alpha = 1$, $E_1\left(-(t/\tau_p)\right) = e^{-t/\tau_p}$ recovers the characteristic exponential relaxation of the standard Rouse model. For $\alpha < 1$, it interpolates between a stretched exponential at short times and a power-law function $t^{-\alpha}/\Gamma(1-\alpha)$ at long times. The relaxation time of the $p$th mode is
\begin{equation}
\tau_p = k_p^{-1/\alpha} = \tau \left(\dfrac{N}{p}\right)^{2/\alpha},
\label{eqn:taup_FRM}
\end{equation}
where $\tau = \left(\zeta/(k \pi^2)\right)^{1/\alpha}$ is the elementary timescale. The longest relaxation or Rouse time corresponds to $p=1$ and is given by $\tau_{R} = \tau N^{2/\alpha}$. For $\alpha=1$, the set of relaxation times $\{\tau_p\}_{p=1}^{N}$ represent the standard Rouse relaxation spectrum. The viscosity $\eta$ is given by
\begin{equation}
\eta = \int_{0}^{\infty} G(t) \, dt = \dfrac{G_0}{N}
  \sum_{p=1}^{N} \int_{0}^{\infty} E_\alpha^2\!\left(-(t/\tau_p)^\alpha\right)\, dt.
 \label{eqn:viscosity_definition}
\end{equation}
For $\alpha < 1$, $E_{\alpha}^2(t \rightarrow \infty) \sim t^{-2\alpha}$, which implies that $\eta$ is finite only when $2\alpha > 1$, or $ \alpha > 1/2$. For $\alpha \leq 0.5$, $\eta$ diverges. 

\begin{figure}
\begin{center}
\includegraphics[scale=0.85]{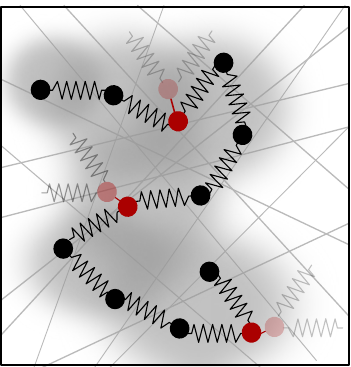}
\caption{Schematic of a tagged linear chain with $N_x = 3$ cross-links or sticky monomers (red) and $N - N_x = 9$ regular (black) monomers immersed in a viscoelastic medium depicted by gray straight lines. Cross-links between sticky beads on the tagged chain and (lighter) chains comprising the environment are explicitly shown to highlight network structure. \label{fig:frm_schematic}}
\end{center}
\end{figure}

\subsection{Fractional Inhomogeneous Rouse Model}

The FRM assumes that the monomer friction $\zeta$ is constant. In the fractional inhomogeneous Rouse model (FIRM), we relax this constraint and allow beads to have different drag coefficients. We also relax the assumption of the continuous limit in the FRM which enables us to model unentangled chains with more confidence; empirically, the number of Kuhn segments per entanglement $N_e \lesssim 30$ for commodity polymer melts.\cite{Everaers2020} Thus, we replace  Eq.~\ref{eqn:GLE_Rouse} with
\begin{equation}
\zeta_n \int_0^t dt'\, K(t-t')\, \dot{\mathbf{r}}_n(t')
  = -\dfrac{\partial U(\mathbf{R})}{\partial \mathbf{r}_n} + \mathbf{f}_n(t), \quad n \in [1, N]
\label{eqn:GLE_firm_indiv}
\end{equation}
where $\zeta_n$ is the drag coefficient of the $n$th bead, $\mathbf{R} = (\mathbf{r}_1, \cdots, \mathbf{r}_N)$ is the set of monomer locations, and $U(\mathbf{R}) = (k/2) \sum_{n=1}^{N-1} (\mathbf{r}_n - \mathbf{r}_{n+1})^2$ is the elastic energy, assuming a constant spring stiffness $k$. We can succinctly represent this system of $N$ equations as
\begin{equation}
\int_0^t dt'\, K(t-t')\, \bGamma \dot{\mathbf{R}}(t') + \mathbf{C} \mathbf{R} = \tilde{\mathbf{F}}(t).
\label{eqn:GLE_firm}
\end{equation}
$\bGamma = \text{diag}(\zeta_1, \cdots, \zeta_N)$ is a diagonal friction matrix. For the homogeneous case, $\bGamma = \zeta \mathbf{I}$, where $\mathbf{I}$ is the identity matrix. $\mathbf{C}$ is a $N \times N$ tridiagonal matrix that encodes polymer connectivity. The off-diagonal elements $C_{i,i \pm 1} = -k$ and the diagonal elements $C_{i,i} = 2k$ for $2 \leq i \leq N-1$. Chain ends modify this pattern for the first and last rows: $C_{1,1} = -C_{1, 2} = -C_{N-1, N} = C_{N, N} = k$. $\tilde{\mathbf{F}} =  (\mathbf{f}_1, \cdots, \mathbf{f}_N)$ is the set of random forces on different beads. It satisfies the bead-level fluctuation-dissipation theorem $\langle \mathbf{f}_{n}(t) \cdot \mathbf{f}_{m}(t')\rangle = 3\zeta_n\,k_BT\, \delta_{nm} K(t-t')$.

\subsubsection{Generalized Normal Modes}

Since the drag is non-uniform, the standard Rouse normal modes (Eq.~\ref{eqn:normal_modes_definition}) cannot be used to diagonalize the spring matrix $\mathbf{C}$ via cosine eigenvectors. Instead, we define generalized normal modes as the solutions of the $\bGamma$-weighted or \textit{generalized eigenvalue problem}:
\begin{equation}
\bC\,\bv_p = \lambda_p\,\bGamma\,\bv_p, \quad p \in [0, N-1]
  \label{eqn:generalized_ev}
\end{equation}
Since $\bC$ is symmetric and positive semi-definite, its eigenvalues are real and nonnegative. Indeed, except for $\lambda_0 = 0$, $\lambda_p > 0$ for $p \geq 1$. The eigenvectors $\bv_p \in \mathbb{R}^N$ follow the $\bGamma$-orthonormality condition:
\begin{equation}
\mathbf{v}_p^T (\bGamma \mathbf{v}_q) = \delta_{pq} \implies \mathbf{V}^T \bGamma \mathbf{V} = \mathbf{I}
\label{eqn:Gamma_normality}
\end{equation}
where the matrix $\mathbf{V} = [\bv_0, \cdots, \bv_p, \cdots ,\bv_{N-1}] \in \mathbb{R}^{N \times N}$ has the eigenvectors stacked as columns. For the homogeneous case with $\bGamma = \zeta \mathbf{I}$, the eigenvectors and eigenvalues for $p \in [0, N-1]$ and $n \in [1, N]$ are given by
\begin{equation}
\begin{gathered}
v_p^{n}  = C_p \cos\left(\dfrac{p \pi (n-0.5)}{N}\right)\\
\lambda_p  = \dfrac{4k}{\zeta} \sin^{2}\left(\dfrac{p \pi}{2N}\right)
\end{gathered}
\label{eqn:ev_homo}
\end{equation}
where $v_p^{n}$ denotes the $n$th element of $\mathbf{v}_p$. The prefactors $C_0 = 1/\sqrt{N\zeta}$ and $C_p = 1/\sqrt{2 N\zeta}$ for $p \geq 1$ ensure that $\mathbf{v}_p \cdot \mathbf{v}_q = \delta_{pq}/\zeta$ follow the $\bGamma$-orthonormality condition.

The generalized eigenvalue problem leads to the following diagonalization
\begin{equation}
\mathbf{V}^T \mathbf{C} \mathbf{V} = \boldsymbol{\Lambda}
\label{eqn:spectral_decomp}
\end{equation}
where $\boldsymbol{\Lambda} = \text{diag}(\lambda_0, \cdots, \lambda_{N-1})$ is a diagonal matrix of eigenvalues. If $\mathbf{X} = [\mathbf{X}_0, \cdots, \mathbf{X}_p, \cdots ,\mathbf{X}_{N-1}]$ is the set of normal modes, then it can be shown using linear algebra that $\mathbf{X} = \mathbf{V}^T \bGamma \mathbf{R}$ and $\mathbf{R} = \mathbf{VX}$. It is important to emphasize that the eigenmodes and generalized normal modes encode the spring and friction matrices and are independent of the memory kernel. 

\subsubsection{Mode Equation of Motion}

We can now decouple the evolution equations for bead positions by projecting onto the generalized normal modes. Multiplying Eq.~\ref{eqn:GLE_firm} with $\mathbf{V}^T$ and using the relations $\mathbf{X} = \mathbf{V}^T \bGamma \mathbf{R}$ and $\mathbf{V}^T \mathbf{C} \mathbf{R} = \boldsymbol{\Lambda} \mathbf{X}$, we obtain
\begin{equation}
\int_0^t dt'\, K(t-t')\, \dot{\mathbf{X}}(t') + \boldsymbol{\Lambda} \mathbf{X}(t) = \mathbf{F}(t).
\label{eqn:GLE_nm_firm}
\end{equation}
where $\mathbf{F} = \mathbf{V}^T \tilde{\mathbf{F}}$. Unlike the tridiagonal matrix $\mathbf{C}$ in Eq.~\ref{eqn:GLE_firm} which couples the motion of the different beads, $\boldsymbol{\Lambda}$ in Eq.~\ref{eqn:GLE_nm_firm} is diagonal and decouples the normal modes.  It allows us to write an independent evolution equation for the $p$th normal mode as,
\begin{equation}
\int_0^t dt'\,K(t-t')\, \dot{\mathbf{X}}_p(t')
  + \lambda_p\,\mathbf{X}_p(t) = \mathbf{F}_p(t).
  \label{eqn:GLE_nmp_firm}
\end{equation}
The projected random force is $\mathbf{F}_p(t) = \mathbf{v}_p^T \tilde{\mathbf{F}} = \sum_{n=1}^{N} v_{p}^{n}
\mathbf{f}_n(t)$. The $\bGamma$-orthonormality that decouples the equations of motion also decouples the noise between modes:
\begin{equation}
  \langle \mathbf{F}_p(t) \cdot \mathbf{F}_q(t')\rangle
 = 3 k_BT\,\delta_{pq}\,K(t-t') 
\label{eqn:random_force_fihr_fdt}
\end{equation}
The equation of motion for the normal modes in FIRM, Eq.~\ref{eqn:GLE_nmp_firm}, is structurally identical to the corresponding equation in FRM, Eq.~\ref{eqn:GLE_Xp}. The extra factor of $C_p^2$ in the prefactor in Eq.~\ref{eqn:fdt_gle_normalmodes} is subsumed in Eq.~\ref{eqn:random_force_fihr_fdt} via $\bGamma$-orthonormality. This implies that $G(t)$ in FIRM also follows Eq.~\ref{eqn:frm_Gt} where $\tau_p = \lambda_p^{-1/\alpha}$ are determined by the nonzero eigenvalues $\{\lambda_p\}_{p=1}^{N-1}$ of the Rouse matrix $\mathbf{C}$.

\subsection{Fractional Sticky Rouse Model}

The sticky Rouse model (SRM), given by Eq.~\ref{eqn:SRM}, is a generalization of the standard Rouse model for transient networks of unentangled polymer chains. It treats beads that form physical bonds or temporary cross-links as sticky beads whose frictional drag coefficient $\zeta_x$ is much greater than the bare monomer friction coefficient $\zeta$. The fractional sticky Rouse model (FSRM) generalizes the SRM for $\alpha \ne 1$. It may be considered as a two-population approximation of FIRM.

It is often assumed that $\tau_x = \tau e^{E_a/RT}$ follows an Arrhenius relation,\cite{Ricarte2021} where $E_a$ is an activation energy for the exchange reaction. In this case, we may approximate the ratio of the drag coefficients $\delta = \zeta_x/\zeta$ for $E_a/RT \gg 1$ as
\begin{equation}
\delta = \dfrac{N_x}{N} \left(\dfrac{\tau_x}{\tau}\right)^{\alpha} = \rho_x e^{\alpha E_a/RT},
\label{eqn:delta}
\end{equation}
where $\rho_x = N_x/N$ is the fraction of sticky beads. Thus, $\tau_x/\tau = (\delta/\rho_x)^{1/\alpha}$, which implies that $\alpha < 1$ amplifies the stickiness of the sluggish beads for fixed $\delta$ and $\rho_x$. For example, $\tau_x/\tau$ increases from  $\delta/\rho_x$ to $(\delta/\rho_x)^2$ as $\alpha$ decreases from 1 to 0.5. Therefore, sub-diffusive dynamics make the chain more sensitive to drag heterogeneity. Although $E_a$ is typically treated as a constant for vitrimer systems, fundamentally neither $\tau_x/\tau$ in the SRM nor $\delta$ in the FIRM are constrained by the Arrhenius relation and potentially can reflect non-Arrhenius processes. We explore this concept in this study.

Regardless of the microscopic mechanism that determines this ratio, the relaxation modulus of the FSRM can be written as:
\begin{equation}
\dfrac{G_\text{FSRM}(t)}{G_0} = \dfrac{1}{N} \left[\sum_{p=1}^{N_x} E_\alpha^2\!\left(- \left(\dfrac{p}{N_x}\right)^2 \left(\dfrac{t}{\tau_x}\right)^\alpha \right) +  \sum_{p=N_x + 1}^{N} E_\alpha^2\!\left(- \left(\dfrac{p}{N}\right)^2 \left(\dfrac{t}{\tau}\right)^\alpha \right) \right]
  \label{eqn:sfrm_Gt}
\end{equation}
When $\alpha = 1$, this is equivalent to the SRM. Based on prior work on this model, we expect the FSRM to be a reasonable approximation to the FIRM when (i) the distribution of sticky beads is uniform or random, (ii) $\tau_x N_x^2 \gg \tau N^2$, and (iii) $N_x \gg 1$. If these conditions are not met, it is preferable to rely on FIRM instead of FSRM.

\section{Results}

The key difference between the fractional (FRM, FIRM, FSRM) and Brownian models (Rouse, IHR, SRM) is the form of the relaxation function. The exponential relaxation in the standard models is replaced by the Mittag-Leffler function in the fractional models. This relatively small mathematical difference has broad ramifications. We begin by exploring the basic properties of the FRM to establish a baseline before considering the effect of sticky beads. Since FRM is a special case of FIRM (figure \ref{fig:models}), we internally use FIRM with $\bGamma = \zeta \mathbf{I}$ in this exploration.

\begin{figure}
\begin{center}
	\includegraphics[scale=0.6]{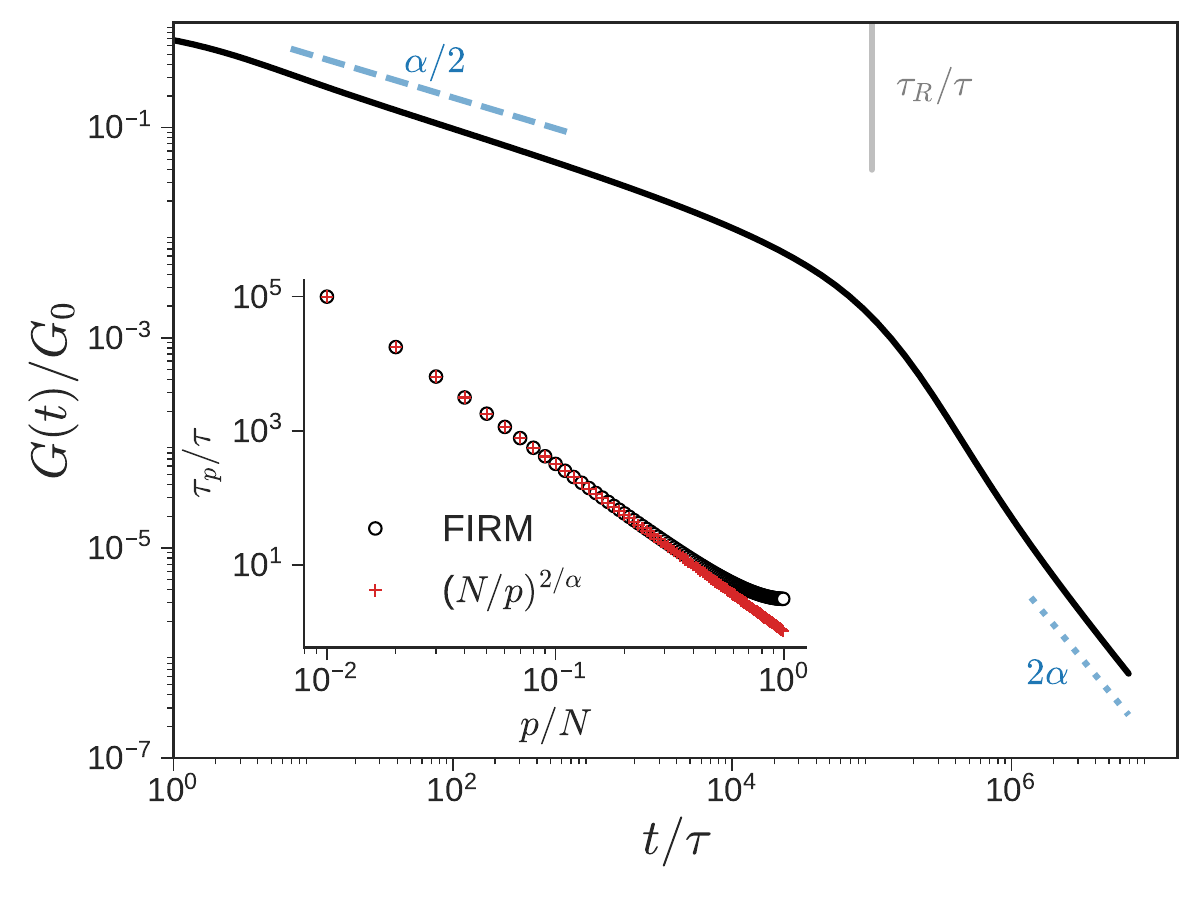}
\end{center}
\caption{The normalized $G(t)$ for a chain with $N= 100$, $\alpha=0.8$, and constant $\zeta$.  $G(t)$ shows two different power-law regimes at intermediate and long times relative to the Rouse relaxation time which is indicated by the vertical gray line from the top axis. The inset compares the true relaxation spectrum (black circles) with the approximation $\tau_p \sim (N/p)^{2/\alpha}$ (red +). The difference in the fast relaxation times does not materially impact $G(t \gtrsim \tau)$. \label{fig:frm}}
\end{figure}

\subsection{Fractional Rouse Model}
 Figure \ref{fig:frm} shows the dimensionless relaxation modulus $G(t)/G_0$ for $N = 100$ and $\alpha = 0.8$. It shows two power-law regimes on either side of the Rouse relaxation time $\tau_R$. For intermediate timescales shorter than the Rouse relaxation time, $t \ll \tau_R$, $G(t) \sim t^{-\alpha/2}$. In this particular example, the observed power-law $t^{-0.4}$ is weaker than the $t^{-0.5}$ dependence of the standard Rouse model. At long times, $t \gg \tau_R$, we observe a stronger power-law $t^{-2\alpha}$ decay. This is qualitatively different from the standard Rouse model which shows exponential decay in the $t \gtrsim \tau_R$ limit.

The inset to figure \ref{fig:frm} shows the spectrum of relaxation times $\tau_p$ obtained using FIRM for this system. For small values of $p$, which correspond to the longest relaxation times, we observe $\tau_p \sim (N/p)^{2/\alpha}$ as expected by the FRM. However, the two spectra diverge at large $p$ due the assumption of $\sin x \approx x$ approximation in the Rouse model.\cite{doipd, Ricarte2021} This difference in fast relaxation times does not cause any noticeable difference in the $G(t)$ obtained using FRM or FIRM for $t \gtrsim \tau$ and $N \gg 1$.

\begin{figure}
\begin{center}
\includegraphics[scale=0.6]{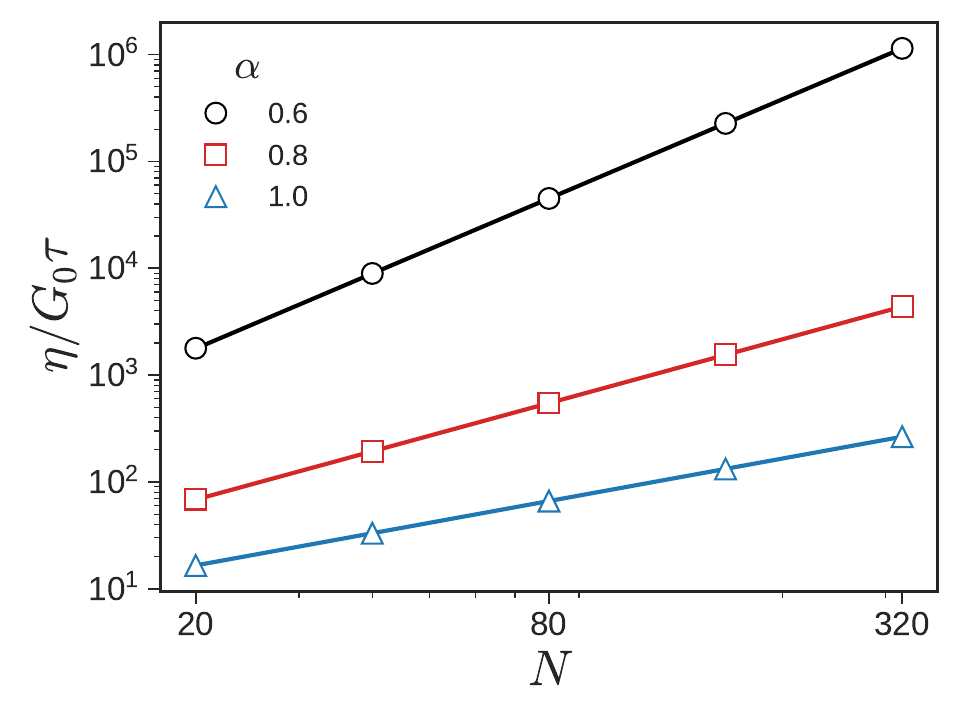}
\end{center}
\caption{Symbols show the numerically computed viscosity of homogenous chains as a function of $N$ at three different values of $\alpha > 1/2$. The viscosity shows a power-law dependence with the chain length $\eta \sim N^{(2/\alpha) - 1}$. The slope of the power-law increases as $\alpha$ decreases. The lines show predictions of the approximate analytical expression $\hat{\eta}(\alpha, N)$ given by Eq.~\ref{eqn:viscosity_approx}.  \label{fig:eta_frm}}
\end{figure}

Figure \ref{fig:eta_frm} shows the viscosity as a function of $N$ and $\alpha$. The numerically computed viscosity $\eta$ in Eq.~\ref{eqn:viscosity_definition} can be approximated with $\hat{\eta}(\alpha, N)$ as:
\begin{equation}
\hat{\eta}(\alpha, N) = \dfrac{G_0 I(\alpha)}{N} \left(\sum_{p=1}^{N} \tau_p\right) \approx G_0 \tau I(\alpha)\, \zeta_{R}(2/\alpha)  N^{(2/\alpha) - 1}
 \label{eqn:viscosity_approx}
\end{equation}
where the sum of relaxation times is approximated with the Riemann zeta function $\zeta_{R}(\cdot)$ assuming $N \gg 1$:
\begin{equation}
\sum_{p=1}^{N} \tau_p \approx \tau_{R} \sum_{p=1}^{\infty}  \dfrac{1}{p^{2/\alpha}} = \tau_{R} \, \zeta_{R}(2/\alpha)
\end{equation}
The subscript $R$ in $\zeta_R(\cdot)$ is used to distinguish the \textit{Riemann} zeta function from the symbol used for the drag coefficient. $\zeta_{R}(2/\alpha)$ decreases monotonically from $\pi^2/6 \approx 1.645$ at $\alpha = 1$ to $\pi^4/90 \approx 1.082$ at $\alpha = 1/2$. The integral $I(\alpha)$ may be approximated with a rational function 
\begin{equation}
I(\alpha) =  \int_{0}^{\infty} E_\alpha^2\!\left(-t^\alpha\right)\, dt \approx \dfrac{0.6488 \alpha^2 - 0.7638 \alpha + 0.3668}{\alpha - 0.5}.
\end{equation}
The approximation was obtained by evaluating the integral numerically and fitting a quadratic formula to the product of $I(\alpha) (\alpha - 0.5)$. The formula is approximately equal to 0.5 at $\alpha = 1$ and diverges as $\alpha$ approaches 0.5 from above.

At each $\alpha$, viscosity increases with chain length as a power-law $\eta \sim N^{2/\alpha - 1}$. This leads to the familiar $\eta \sim N$ dependence for the standard Rouse model. Interestingly, as $\alpha$ approaches 0.5 from above, the slope increases and asymptotically matches the pure reptation limit of $\eta \sim N^{3}$.\cite{doipd} The overlap between the numerically computed $\eta$ (symbols) and the analytical expression for $\hat{\eta}$ (lines) in figure \ref{fig:eta_frm} validates the approximations for $I(\alpha)$ and the sum of relaxation times with the Riemann zeta function. At a particular value of $N$, viscosity increases as $\alpha$ decreases before diverging at $\alpha = 0.5$. This can be understood using the expression for $\hat{\eta}$. The substantial increase in $\tau_R$ and $I(\alpha)$ more than makes up for the modest decrease in $\zeta_R(2/\alpha)$ as $\alpha$ approaches 0.5.

The transition between the two power-laws observed in figure \ref{fig:frm} is reminiscent of the fractional Maxwell model (FMM), which is a phenomenological model consisting of two springpots.\cite{Friedrich1991} As expected, the shape of the FMM roughly matches the shape of $G(t)$ determined by FRM for $\alpha < 0.5$. However, despite the same terminal slopes (see figure \ref{fig:fmm_fit} in the Appendix), the shapes of the FMM and FRM differ --- the FRM cannot be quantitatively described by a FMM.

\subsection{FIRM for Transient Unentangled Polymer Networks}


\begin{figure}
\begin{center}
\includegraphics[scale=0.7]{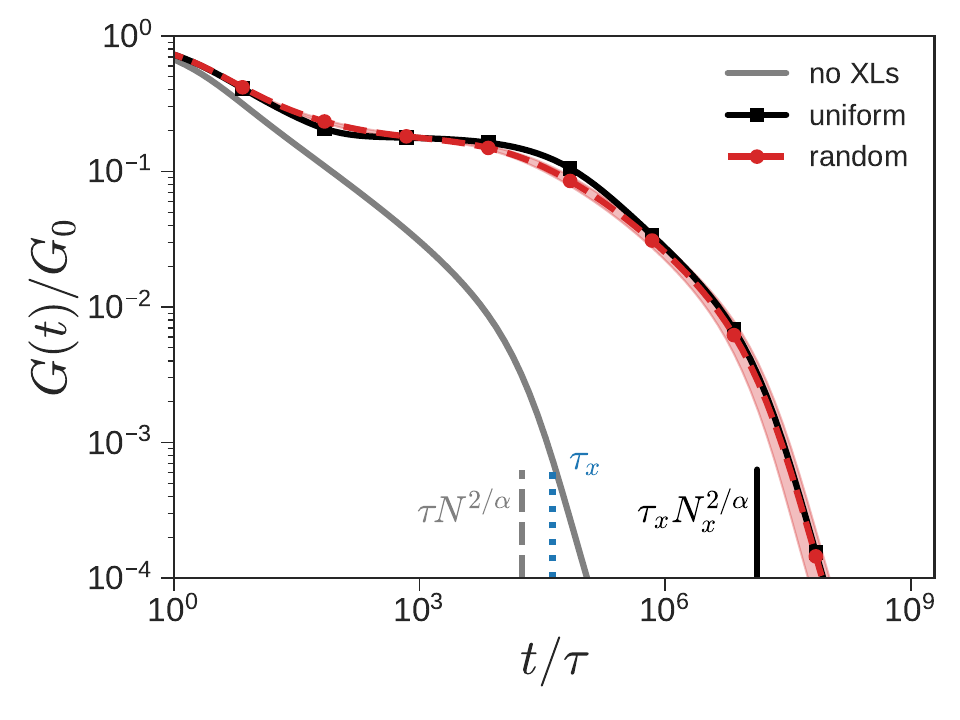}
\end{center}
\caption{FIRM predictions of the relaxation modulus for uniformly (squares) or randomly (circles) distributed cross-links are quite similar. For randomly distributed cross-links, the shaded region shows the standard deviation using 50 replicas. The parameters used were $N=50$, $\rho_x = 0.2$, $\alpha=0.8$, and $\delta = 10^{3}$. The solid gray line shows the relaxation of a bare chain without any cross-links ($\rho_x = 0$). The Rouse relaxation time of the bare chain, the lifetime of a cross-link, and terminal relaxation time are indicated by vertical stems on the bottom axis.
 \label{fig:compUniRand}}
\end{figure}

We now explore the properties of FIRM that can be used to model transient polymer networks using non-uniform frictional coefficients. The gray line in figure \ref{fig:compUniRand} establishes as a baseline: it shows the relaxation modulus of a chain with $N = 50$ monomers and $\alpha = 0.8$ without any cross-links. We observe the transition in the power-law exponent from $\alpha/2 = 0.4$ to $2\alpha = 1.6$ around the Rouse relaxation time $\tau_R = \tau N^{2/\alpha}$, which is indicated by a dashed gray stem on the horizontal axis.

For chains with cross-links, we set cross-link density $\rho_x = 0.2$ and $\delta = \zeta_x/\zeta = 10^3$. The solid line with squares shows the relaxation modulus when the $N_x = \rho_x N = 10$ cross-links are uniformly distributed along the chain backbone. The $G(t)$ of the cross-linked chain deviates from the bare chain. We observe a plateau in $G(t)/G_0 = \rho_x$ which arises from the persistence of the cross-links. This plateau is followed by the two familiar power-law regimes ($\alpha/2$ and $2 \alpha$). The first power-law regime starts around $t \approx \tau_x$ corresponding to the lifetime of a cross-link. This transitions to the stronger power-law near the longest relaxation time, $\tau_x N_x^{2/\alpha}$, of the cross-linked chain.

The distribution of the sticky monomers can affect stress relaxation. However, the mean $G(t)$ of chains in which cross-links are randomly distributed (with average $N_x = \rho_x N$) --- shown by the  circles and dashed line in figure \ref{fig:compUniRand} --- is quite similar to the response of chains where cross-links are uniformly distributed. Here, the mean $G(t)$ is calculated by averaging over an ensemble of 50 independent chains. The shaded red region shows the standard deviation computed from these replicas. Comparison of uniformly and randomly distributed stickers at different values of $\rho_x$, $N_x$, and $\delta$ reveal the universality of this correspondence. This conclusion is identical to that obtained with the IHR model.\cite{Ricarte2021} Thus, all subsequent results of the FIRM are reported for uniformly distributed cross-links unless mentioned otherwise.

\begin{figure}
\begin{center}
\includegraphics[scale=0.6]{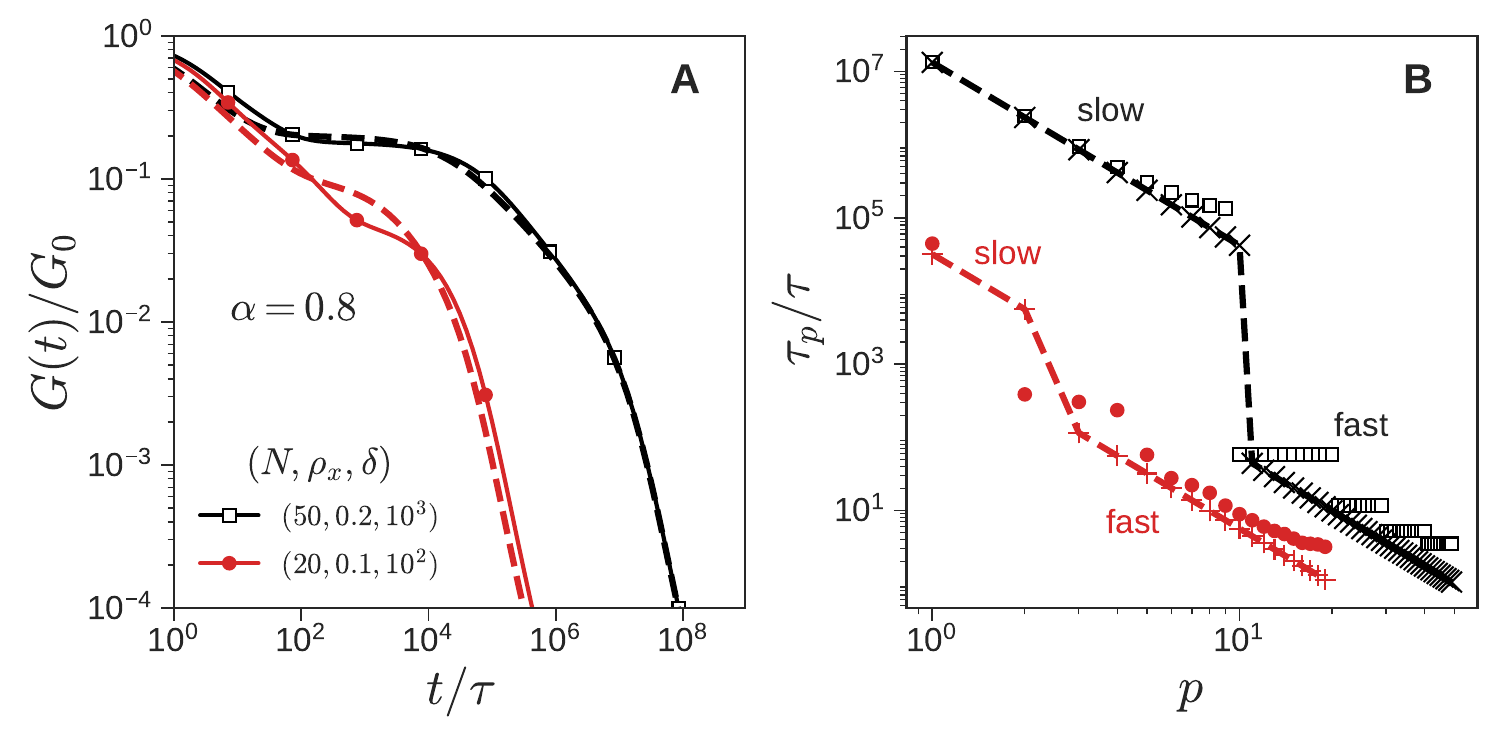}
\end{center}
\caption{(A) The agreement between FIRM (symbols with solid lines) and FSRM (dashed lines) is superior when $\delta = 10^{3}$ and $N_x = \rho_x N = 10$ are both large (squares). It worsens as $\delta = 10^{2}$ and $N_x = 2$ decrease (circles). (B) The spectrum of relaxation times separates into distinct slow and fast modes when $\delta$ and $N_x$ are large as shown by the squares (FIRM) and crosses connected by a dashed line (FSRM). This separation is compromised as $\delta$ and $N_x$ decrease as shown by the circles (FIRM) and dashed line connecting + signs (FSRM). \label{fig:sfrm_validity}}
\end{figure}

Next, we compare results of the FSRM, which provides an analytical expression for $G(t)$, with that of the FIRM in which the relaxation times have to be evaluated numerically. Figure \ref{fig:sfrm_validity}A contrasts these two models at a fixed value of $\alpha = 0.8$. FIRM predictions are identified with symbols joined by a solid line, while the FSRM predictions are shown using dashed lines. The corresponding spectrum of relaxation times are shown in figure \ref{fig:sfrm_validity}B. For the FSRM, symbols ($\times$) and (+) are added to the dashed lines to mark the discrete values of $\tau_p$. For clarity, the solid lines connecting the relaxation times corresponding to FIRM are dropped.

When $N = 50$ and $\rho_x = 0.2$, the number of cross-links $N_x = \rho_x N = 10$ which is sufficiently large. When combined with a strong contrast in bead mobilities ($\delta = 10^3$), the assumptions embedded in the FSRM are satisfied. In this scenario, there is general agreement in the $G(t)$ obtained using the two models. The spectrum of relaxation times is shown in figure \ref{fig:sfrm_validity}B. First, there is clear separation in the slower and faster relaxation times. Furthermore, there is close correspondence in the slower relaxation times ($p \lesssim 5$), which explains the agreement in $G(t)$ for $t \gtrsim 100\tau$. The spectrum of the faster relaxation modes is interesting: while $\tau_p \sim p^{-2/\alpha}$ in the FSRM, the numerically computed relaxation times in the FIRM form clusters in which $\tau_p$ is approximately constant for a series of adjacent modes. This blocky pattern arises because the mobility of regular beads sandwiched between sticky beads is constrained by the mobility of the sticky beads. Consequently, the $G(t)$ curves obtained using the FIRM and FSRM do not overlap cleanly at short times.

Filled circles repeat the calculation with $N = 20$, $N_x = 2$ and $\delta = 10^2$, which is typical of many experimental systems. In this regime, we observe discrepancies between FSRM and FIRM. For example, the shoulder in the $G(t)$ computed using FIRM is softer than that obtained using FSRM. The relaxation spectrum provides some explanation for this observation. Although the slowest relaxation time ($p=1$) is comparable, there is a large difference in $\tau_2$ because of the interaction of the exchange and segmental relaxation modes. Again, this observation is similar to the IHR: FSRM offers a decent analytical approximation to the FIRM when $N_x \gg 1$, $\delta \gg 1$, and the cross-links are uniformly or randomly distributed.

\section{Discussion}

\subsection{Experimental Benchmarks}

We consider previously reported linear viscoelastic measurements on a polystyrene vitrimer with imine cross-links labeled PS-v-8-1.2.\cite{Barzycki2025} It was synthesized from precursor chains with a number-average molar mass of 8 kDa, which is sufficiently below the entanglement molecular weight and corresponds to the first number in the sample label. The second number in the label corresponds to the molar ratio of the diamine cross-linker and the aldehyde pendant groups, indicating an excess of the cross-linkers.

\begin{figure}
\begin{center}
\begin{tabular}{cc}
\includegraphics[scale=0.5]{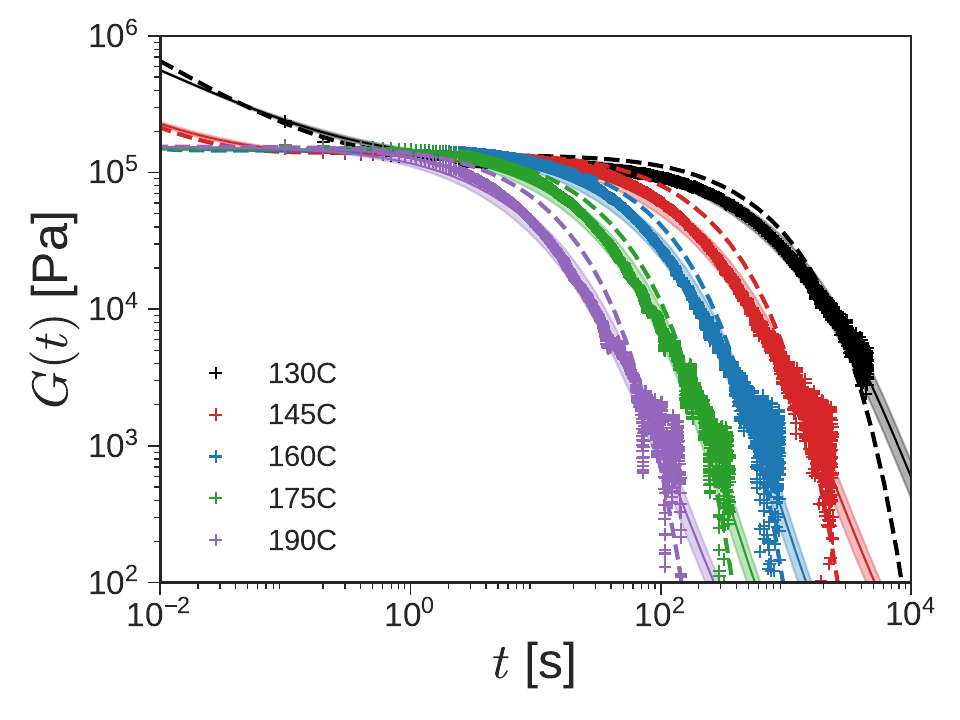} & \includegraphics[scale=0.5]{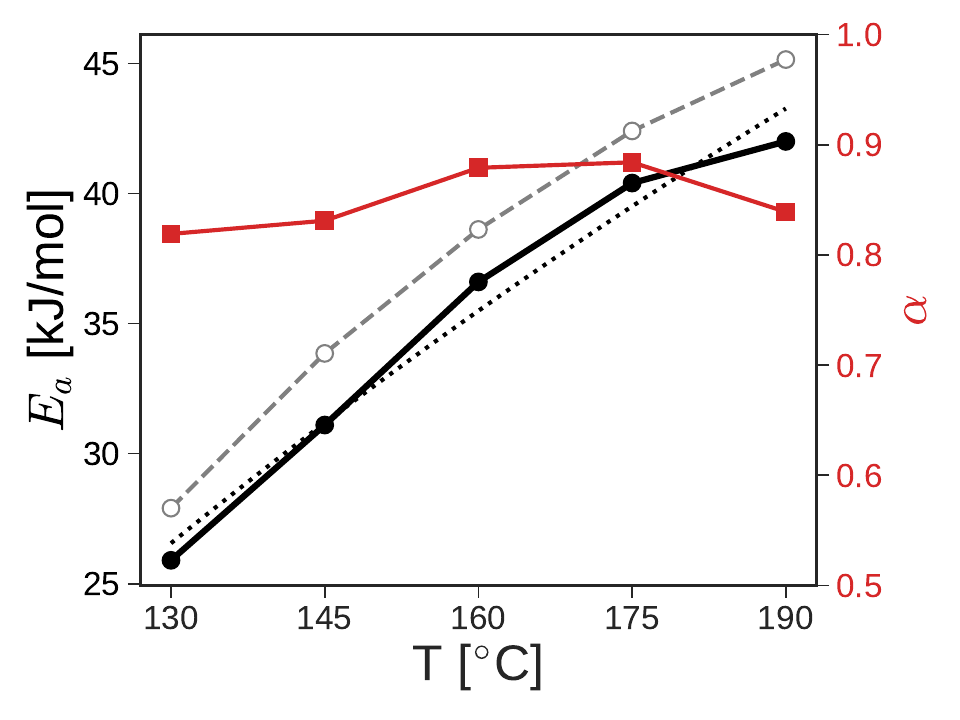}\\
(A) stress relaxation & (B) fitted parameters
\end{tabular}
\end{center}
\caption{(A) Fits of the FIRM  to experimental stress relaxation curves (+) for unentangled polystyrene vitrimers at different temperatures. Solid lines show fits with $\alpha$ and $E_a$ as free parameters. Shaded regions indicate uncertainty using 100 independent replicas. Dashed lines show fits of FIRM with $\alpha =1$ held fixed, and $E_a$ and the only free parameter. (B) Filled symbols show the best-fitting parameters $\alpha$ (squares) and $E_a$ (circles) on the right and left vertical axes, respectively, for each dataset. The dotted line shows an approximate fit $E_a(T) = 155.5 - 5.1952 \times 10^{4}/T$. Open symbols connected by the dashed line shows the best fitting $E_a$ with $\alpha=1$ held constant. \label{fig:PS_Gt_fit}}
\end{figure}

Stress relaxation experiments were performed at five temperatures between 130 -- 190 $^{\circ}$C shown by symbols in figure \ref{fig:PS_Gt_fit}A. Higher temperatures promote segmental mobility and bond-exchange, leading to faster relaxation. Thus, the $G(t)$ at 190 $^{\circ}$C decays more rapidly than at 130 $^{\circ}$C in figure \ref{fig:PS_Gt_fit}A.

The molar mass of a polystyrene Kuhn segment is 720 g/mol.\cite{RubinsteinPP} Thus, the number of Kuhn segments per chain $N \approx 11$. The effective number of cross-links per chain was found to be 2.1, implying a cross-link density $\rho_x = 0.189$. Bare-chain parameters were obtained from a well-studied unentangled polystyrene melt (PS13K) at 104.7 $^{\circ}$C.\cite{Santangelo2001} The molar mass dependence implied by the standard Rouse model,\cite{rouse53} and the temperature dependence implied by the WLF equation \cite{Williams1955} 
\begin{equation}
\log_{10} a_\text{WLF}(T) = \dfrac{-C_1 ( T - T_\text{ref} )}{C_2 + T - T_\text{ref}}
\label{eqn:a_wlf}
\end{equation}
with $C_1 = 7.57$ and $C_2 = 115.7$ K were used to shift the parameters to determine the appropriate parameters for PS-v-8-1.2. After disentangling the contribution of the glassy modes,\cite{Poh2020} the Rouse modulus $G_0 \approx 1.54$ MPa. The elementary Rouse time $\tau$ is adjusted to 7.2 ms to account for the nearly 15 $^{\circ}$C gap in the glass transition temperatures of PS13K (87.6 $^{\circ}$C) and PS-v-8-1.2 (103 $^{\circ}$C).

The only unknown parameters are $\alpha$ and $\delta$, or equivalently, $E_a$ (Eq.~\ref{eqn:delta}). We determined $\alpha$ and $E_a$ by fitting FIRM to each experimental dataset. To systematically evaluate the influence of each parameter, we conduct the fitting by treating (i) both $E_a$ and $\alpha$, (ii) only $E_a$, or (iii) only $\alpha$ as fitted parameters.

Figure \ref{fig:PS_Gt_fit}A shows the fits when both $E_a$ and $\alpha$ are fitted parameters, while figure \ref{fig:PS_Gt_fit}B displays the best-fitting values of $\alpha$ and $E_a$. The value of $\alpha$ is confined to a relatively narrow range between 0.85 $\pm $ 0.03, while $E_a$ shows a systematic increase from approximately 26 kJ/mol at 130 $^{\circ}$C to 42 kJ/mol at 190 $^{\circ}$C.  This dependence can be approximated using
\begin{equation}
E_a(T) = 155.5 - \dfrac{5.1952 \times 10^{4}}{T}
\label{eqn:Ea_fit}
\end{equation}
shown by the dotted line in figure \ref{fig:PS_Gt_fit}B. The units of $E_a$ and $T$ in this correlation are kJ/mol and K, respectively. The increase in $E_a$ with temperature suggests that the slow relaxation regime of the PS vitrimer follows an apparent non-Arrhenius temperature dependence.

Figure \ref{fig:PS_Gt_fit}A also depicts the fits in which $E_a$ is a fitted parameter and $\alpha$ is held to a constant value of 1 – a case equivalent to the IHR model. The resulting $G(t)$ fits are noticeably poorer and predict an abrupt, exponential terminal relaxation. The best-fitting values of $E_a$ for this case, shown by the dashed line in figure \ref{fig:PS_Gt_fit}B, follow a non-Arrhenius trend similar to the unconstrained fit with values that are about 2--3 kJ/mol higher. The poor quality of the fits demonstrates that using $\alpha < 1$ is important for capturing the shape of the experimentally observed $G(t)$ curves, especially at long times.

Figure \ref{fig:EaConstant} in the appendix displays the fits that have a constant value of $E_a$ = 35 kJ/mol and an adjustable $\alpha$. The $G(t)$ fits are quite poor in this case. In general, holding $E_a$ fixed, instead of increasing with $T$, leads to a stronger temperature dependence for stress relaxation than is observed in experiments. In other words, if $E_a$ is held constant, $G(t)$ calculated by FIRM exhibits a greater spread than shown in figure \ref{fig:PS_Gt_fit}A: relaxation of $G(t)$ at 190 $^{\circ}$C is accelerated (shift to the left), while that at 130 $^{\circ}$C is decelerated (shift to the right). This is not surprising because it has been empirically demonstrated that bond-exchange in polystyrene vitrimers with diamine cross-linkers is not purely reaction-limited as assumed by setting $\tau_x = \tau e^{E_a/RT}$,\cite{Ricarte2021, Barzycki2025} but also depends on the diffusivity of the cross-linker.\cite{Barzycki2026} Thus, the increase in $E_a$ shown in figure \ref{fig:PS_Gt_fit}B indirectly accounts for this process.

\begin{figure}
\begin{center}
\includegraphics[scale=0.6]{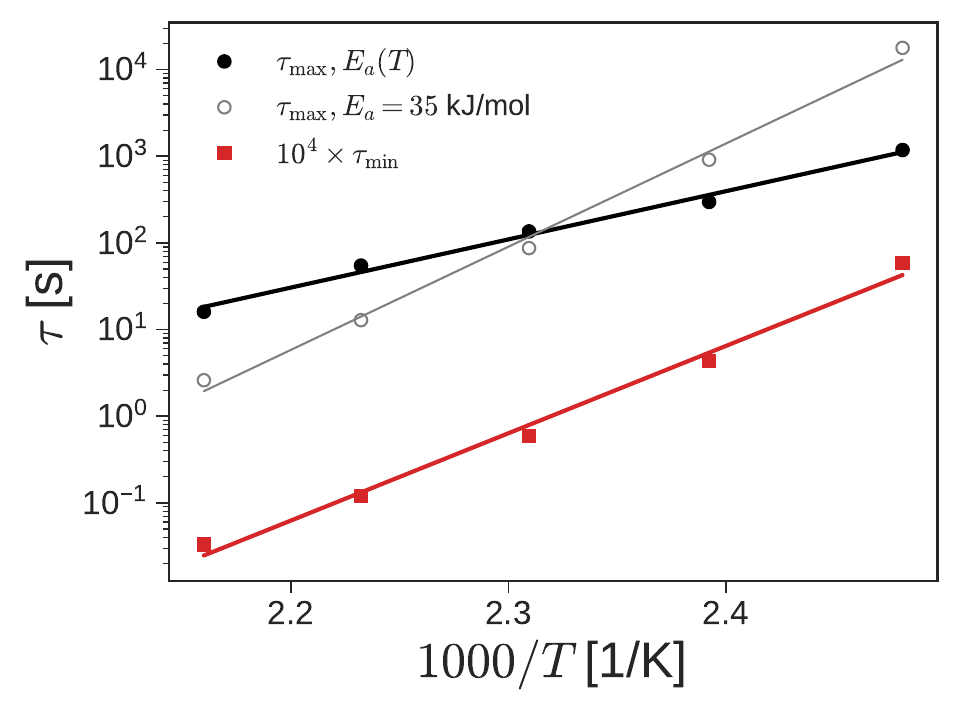}
\end{center}
\caption{The maximum $\tau_{\max}$ (circles) and minimum $\tau_{\min}$ (squares) relaxation times predicted by FIRM as a function of inverse temperature for a system with $N$ = 10, $\rho_x$ = 0.2, and $\alpha$ = 0.85. Filled and open circles show $\tau_{\max}$ when $E_a(T)$ given by Eq. \ref{eqn:Ea_fit} and $E_a$ = 35 kJ/mol, respectively. $\tau_{\min}$ is independent of $E_a$; data are shifted up by four orders of magnitude for visualization. Lines are linear fits of the $\log \tau$ versus $1/T$ data.
\label{fig:timescales}}
\end{figure}

Figure \ref{fig:timescales} further isolates this phenomenon by investigating the temperature dependence of the longest ($\tau_{\max}$) and shortest ($\tau_{\min}$) relaxation times for different choices of $E_a(T)$. For this investigation, we simplify the polystyrene vitrimer system, by 
\begin{enumerate}
\item [(i)] setting $N = 10$, $\rho_x = 0.2$, and $\alpha = 0.85$;
\item [(ii)] assuming a uniform distribution of cross-links to avoid dealing with ensembles;
\item [(iii)] considering two different choices for $E_a(T)$: an increase with $T$ shown by the filled symbols in figure \ref{fig:PS_Gt_fit}B and a constant value of $E_a$ = 35 kJ/mol.
\end{enumerate}
From the spectrum of relaxation times produced by FIRM, we obtained $\tau_{\min}$ and $\tau_{\max}$ at the five different temperatures between 130 -- 190 $^{\circ}$C. The variation of $\tau_{\min}$ on temperature does not depend on the functional form or magnitude of $E_a$. This is expected because $\tau_{\min}$ corresponds to segmental relaxation time. Its temperature dependence is governed by WLF activation energy,
\begin{equation}
\dfrac{E_\text{WLF}(T)}{R} =  2.303 \dfrac{\partial \log_{10} a_T}{\partial(1/T)}= \dfrac{2.303 C_1 C_2 T^2}{(C_2 + T - T_\text{ref})^2}.
\label{eqn:Ewlf}
\end{equation}
$E_\text{WLF}$ depends on the arbitrary choice of $T_\text{ref}$ unless $T \rightarrow \infty$. Therefore, the magnitudes of $E_\text{WLF}$ reported in the following discussion, where we set $T_\text{ref} = 418$ K, should be interpreted with caution.

With this caveat, $E_\text{WLF}$ varies from approximately 270 kJ/mol at 130 $^{\circ}$C to 140 kJ/mol at 190 $^{\circ}$C. Nevertheless, the curvature in figure \ref{fig:timescales} is small enough to warrant linear regression. Assuming $\tau_{\min} \sim e^{E_{\min}/RT}$, we can estimate $E_{\min} \approx $ 193 kJ/mol from the slope of $\log \tau_{\min}$ versus $1/T$. This value is approximately midway between the $E_\text{WLF}$ estimated via Eq. \ref{eqn:Ewlf} at the extremes of the temperature range probed.

In contrast the variation of $\tau_{\max}$ depends sensitively on the choice of $E_a(T)$. When $E_a$ is assumed to be Arrhenius with a constant value of 35 kJ/mol, $E_{\max}$ is 228 kJ/mol, which is 35 kJ/mol larger than $E_{\min}$. Thus, $E_{\max} = E_{\min} + E_a$ as anticipated by the relation $\tau_{\max} = \tau_{\min} e^{E_a/RT}$. When $E_a$ is assumed to be non-Arrhenius with a value that increases with temperature, it decreases the slope of $\log \tau_{\max}$ on $1/T$ and tilts the curve downwards. This slows down dynamics at high temperatures by increasing $\tau_{\max}$ at small values of $1/T$, while speeding the dynamics at low temperatures by decreasing $\tau_{\max}$ at large values of $1/T$. The effective activation energy corresponding to $\tau_{\max}$ is 106 kJ/mol, which is lower than $E_\text{WLF}$ and comparable with the 117 $\pm$ 5 kJ/mol obtained empirically via time-temperature superposition.\cite{Barzycki2025} This is evident from the slopes of the $\tau_{\max}$ and $\tau_{\min}$ curves in figure \ref{fig:timescales}.

If $\tau_{\max} \sim \tau_{\min} e^{E_a(T)/RT}$, where $\tau_{\min} \sim e^{E_{\min}/RT}$, we can write,
\begin{equation}
\dfrac{E_{\max}}{T} = \dfrac{E_{\min}}{T} + \dfrac{E_{a}(T)}{T}
\label{eqn:energy_sum}
\end{equation}
In the case of an Arrhenius $E_a$ that is constant, this simplifies to the familiar expression $E_{\max} = E_{\min} + E_a$. In the case of a non-Arrhenius $E_a$ that increases with temperature as $E_a(T) = a - b/T$ (see Eq. \ref{eqn:Ea_fit}), we obtain
\begin{equation}
E_{\max} = E_{\min} + \dfrac{d E_a(T)}{d (1/T)} = E_{\min} + a - \dfrac{2b}{T}
\end{equation}
For PS-v-8-1.2, $a = 155.5$ kJ/mol and $b = 5.1952 \times 10^{4}$ kJ.K/mol. Using the median value of temperature $T$ = 433 K, this expression implies $E_{\max}$ = 108 kJ/mol which is quite close to the value of 106 kJ/mol extracted from the slope of $\tau_{\max}$.

In summary, FIRM may evaluate how $\alpha$ and $E_a$ vary under different constraints. $\alpha < 1$ is required to capture the correct $G(t)$ curvature. $E_a$ may be treated as independent or dependent on temperature, accommodating both Arrhenius and non-Arrhenius behavior.


\subsection{Application of FIRM to Other Dynamic Properties}

The dynamics of transient polymer networks can be complex. Often, multiple characterization techniques are required to tease apart concurrent processes that occur at different time and length scales. An advantage of a microscopic model like FIRM is that it is not restricted to linear viscoelasticity. One can readily extract other dynamic properties such as the autocorrelation of the end-to-end vector, $\mathbf{R}_e = \mathbf{r}_{N} -\mathbf{r}_{1}$, which can be measured via dielectric relaxation for certain chemistries or the mean-squared displacement (MSD) of cross-links or the polymer center-of-mass $\rho_\text{cm}$, which can be measured via neutron scattering or spin echo measurements.\cite{Oeser1988, Richter1989, Gold2017} This is valuable for transient  networks where no single characterization technique provides a full picture.

Incorporating FIRM into the analysis of dielectric relaxation data offers a potentially powerful pathway to evaluate transient networks. Permanent dipoles of type-A polymers such as polyisoprene are oriented parallel to the polymer backbone, making the net macroscopic dipole moment proportional to the polymer end-to-end vector $\mathbf{R}_{e} = \mathbf{r}_N - \mathbf{r}_1$.\cite{Stockmayer1967, Watanabe2001, Watanabe2002} Cross-links containing imines or vinylogous urethanes also possess localized dipole moments, which introduce an additional contribution that tracks the rearrangement and life-cycle of dynamic bonds.\cite{Tress2019, Ge2020, Ge2023, Shanbhag2023} Dielectric spectroscopy can therefore be used to tease apart segmental and chain dynamics in vitrimer systems.\cite{Arbe2023, Alegria2024} 

\begin{figure}
\begin{center}
\begin{tabular}{cc}
\includegraphics[scale=0.5]{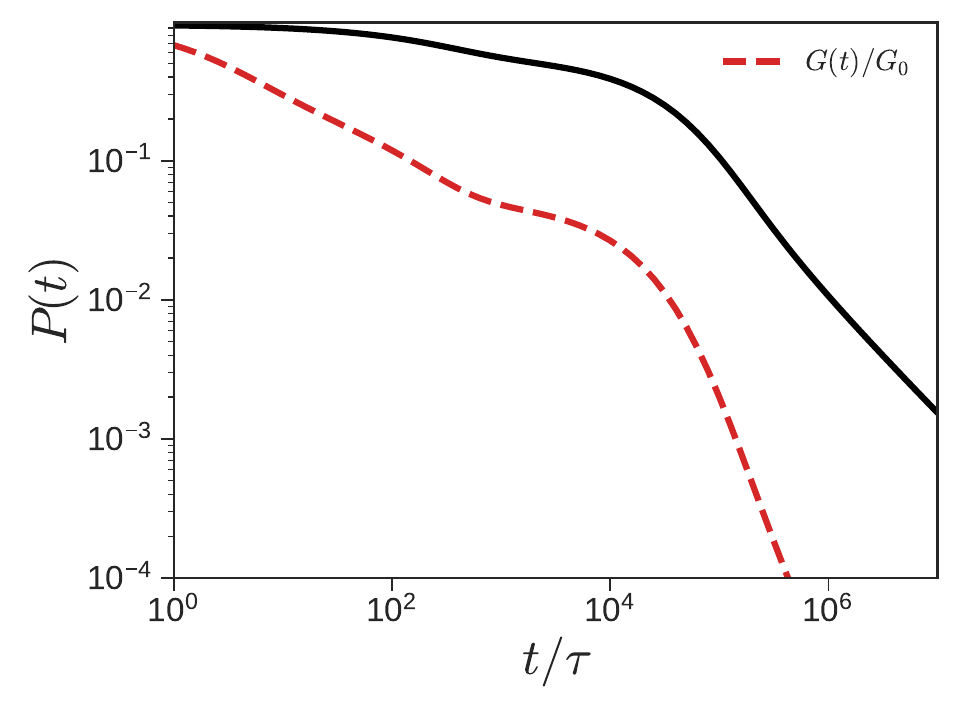} & \includegraphics[scale=0.5]{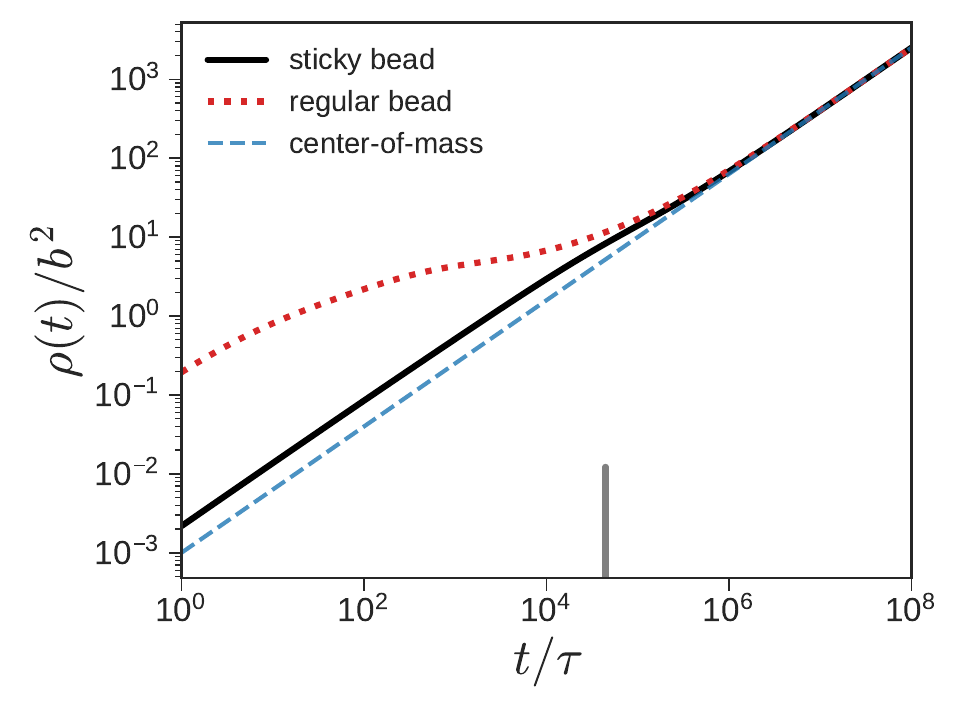}\\
(A) end-to-end vector & (B) mean-squared displacement
\end{tabular}
\end{center}
\caption{(A) The autocorrelation of the end-to-end vector $P(t)$ for the $N=20$ and $\rho_x = 0.1$ chain, previously considered in figure \ref{fig:sfrm_validity}. The dashed line shows the normalized stress relaxation modulus for comparison. (B) The mean-squared displacement of the regular (dotted) and sticky (solid) beads.  The dashed line shows the MSD of the chain center-of-mass, which scales as $t^{\alpha}$. The vertical line on the horizontal axis marks the longest relaxation time. \label{fig:dynaProp}}
\end{figure}

Once the eigenvalues ($\lambda_p$) and eigenvectors ($\mathbf{v}_p$) corresponding to the generalized normal modes are computed, all dynamic properties can be evaluated. For example, starting from $\mathbf{r}_n(t) = \sum v_p^{n} \mathbf{X}_p(t)$, the autocorrelation function of the end-to-end vector in FIRM is found to be
\begin{equation}
P(t) = \dfrac{\ave{\mathbf{R}_{e}(t) \cdot \mathbf{R}_{e}(0)}}{\ave{R_{e}^2(0)}} = \dfrac{3 k_B T}{Nb^2} \sum_{p=1}^{N-1}  \dfrac{c_p^2}{\lambda_p} E_{\alpha} \left(-(t/\tau_p)^{\alpha} \right)
\label{eqn:Ree_auto}
\end{equation}
where $c_p = v_p^N - v_p^1$ is the projection coefficient of the $p$th mode and $\tau_p^{\alpha} = \lambda_p^{-1}$. Figure \ref{fig:dynaProp}A shows the relatively slower decay of $P(t)$ vis-a-vis $G(t)$ for a sticky polymer with $N = 20$, $\rho_x = 0.1$, $\alpha = 0.8$, and $\delta = 100$ with uniformly distributed cross-links. The difference between $P(t)$ and $G(t)$ sheds differential insights into the relaxation process, which can be harnessed to test and refine microscopic theories.\cite{Watanabe2000, Shanbhag2001}

FIRM can also be used to compute the MSD of the $n$th bead $\rho_n(t) = \ave{\left( \mathbf{r}_n(t) - \mathbf{r}_n(0) \right)^2}$, given by
\begin{equation}
\rho_{n}(t) = \rho_{\text{cm}}(t) + \sum_{p=1}^{N-1} \left(v_p^n \right)^2 \cdot \dfrac{6 k_B T}{\lambda_p} \left[1 - E_{\alpha} \left(-(t/\tau_p)^{\alpha} \right) \right]
\label{eqn:msd_n}
\end{equation}
where  $\rho_{\text{cm}}$ is the MSD of the center-of-mass given by
\begin{equation}
\rho_{\text{cm}}(t) = \dfrac{6 k_B T v_0^2}{\Gamma(1 + \alpha)} t^{\alpha}.
\end{equation}
$\mathbf{v}_0 = v_0 \mathbf{1}$ is the constant eigenvector corresponding to the zeroth eigenmode $\mathbf{X}_0(t)$. Figure \ref{fig:dynaProp}B shows that the  center-of-mass MSD, $\rho_{\text{cm}}(t) \propto t^{\alpha}$, follows a power-law. The dotted and solid lines in the figure show the average MSD of the regular and sticky beads, respectively. At short timescales, regular beads are more mobile than the sticky beads. However, at timescales greater than the longest relaxation time, the MSDs of all beads converge to $\rho_{\text{cm}}(t)$.

\medskip

Experimentally, dielectric and stress relaxation are often probed in the frequency-domain using oscillatory perturbations. The complex dielectric permittivity $\epsilon^{*}(\omega)$ is related to $P(t)$ via a Fourier transform
\begin{equation}
\dfrac{\epsilon^{*}(\omega) -\epsilon_{\infty}}{\epsilon_0 - \epsilon_{\infty}} = 1 - i \omega \int_{0}^{\infty} P(t)\, e^{-i\omega t} dt
\label{eqn:defn_cmplx_permittivity}
\end{equation}
where $\epsilon_0$ and $\epsilon_{\infty}$ are the static (zero-frequency) and high-frequency permittivity limits, respectively. The integral defines the one-sided Fourier transform of a real causal function. The Fourier transform of the Mittag-Leffler function may be expressed via,
\begin{equation}
i \omega \int_{0}^{\infty} E_\alpha\left(-(t/\tau)^{\alpha}\right) \,e^{-i\omega t}\, dt = \dfrac{ (i\omega \tau)^{\alpha}}{1 + (i\omega \tau)^{\alpha}}
\label{eqn:ft_ML}
\end{equation}
Therefore, Eqs. \ref{eqn:Ree_auto}, \ref{eqn:defn_cmplx_permittivity}, and \ref{eqn:ft_ML} imply
\begin{equation}
\dfrac{\epsilon^{*}(\omega) -\epsilon_{\infty}}{\epsilon_0 - \epsilon_{\infty}} =
\dfrac{3 k_B T}{Nb^2} \sum_{p=1}^{N-1}  \dfrac{c_p^2}{\lambda_p} \, \dfrac{1}{1 + (i\omega \tau_p)^{\alpha}}.
\label{eqn:epsilon_star}
\end{equation}
Thus the dielectric response of the FIRM can be represented by a linear combination of elements that follow the Cole-Cole model.\cite{kremer2002broadband} Relative to the classic Debye model ($\alpha = 1$), the $\alpha$ parameter in the Cole-Cole model broadens the loss peak symmetrically on a log-frequency axis.

\medskip

\begin{figure}
\centering
\includegraphics[scale=0.6]{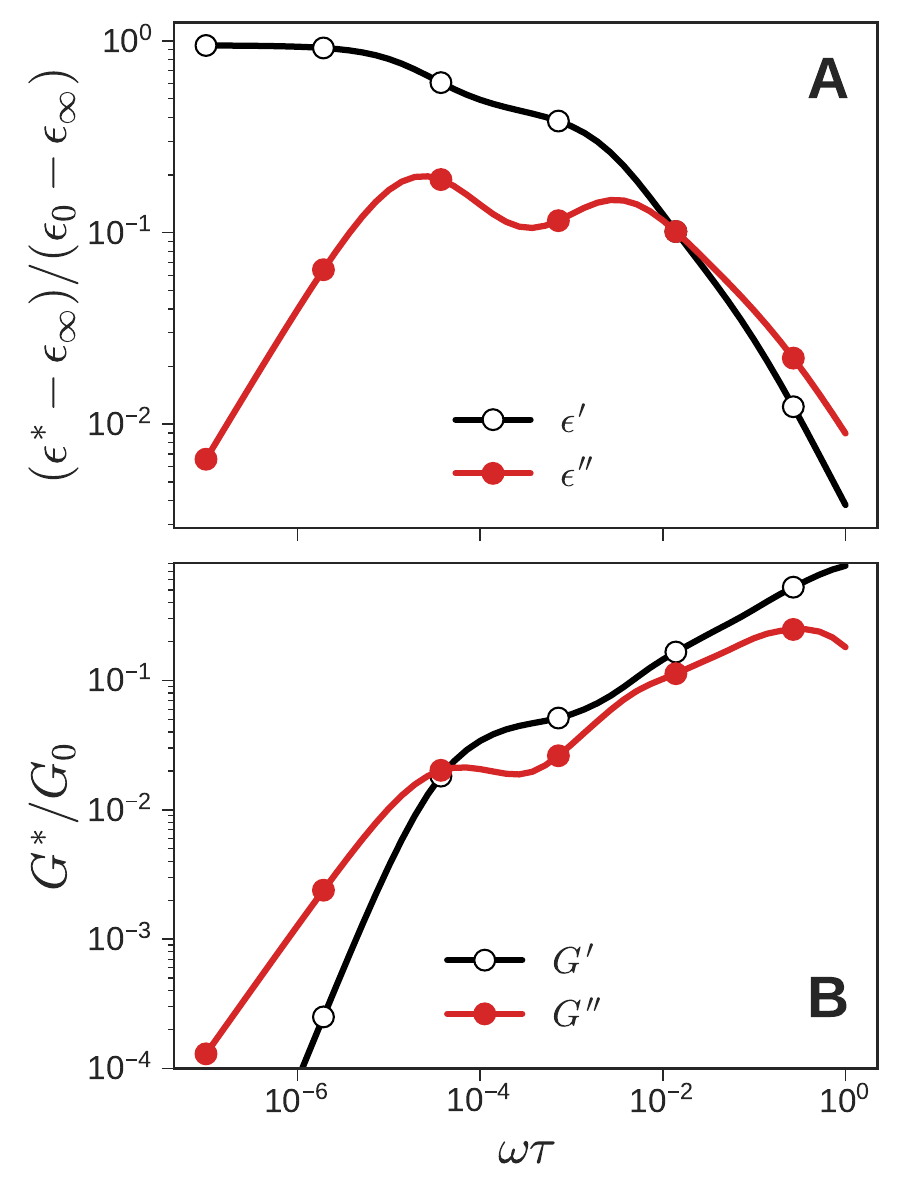}
\caption{Frequency-domain representations of the relaxation functions $P(t)$ and $G(t)$ in figure \ref{fig:dynaProp}A. (A) The dielectric storage and loss functions are the real and imaginary parts of $\epsilon^{*}$ evaluated via Eq. \ref{eqn:epsilon_star}. (B) The storage and loss modulus are the real and imaginary parts of $G^{*}$. They are calculated by fitting $G(t)$ to a relaxation spectrum.
\label{fig:freqProp}}
\end{figure}

Unlike $\epsilon^{*}(\omega)$, we cannot write a closed-form expression for the complex modulus
\begin{equation}
G^{*}(\omega) = i \omega \int_{0}^{\infty} G(t)\, e^{-i\omega t} dt
\label{eqn:defn_cmplx_modulus}
\end{equation}
because $G(t)$ combines terms containing squares of the Mittag-Leffler function, $E_\alpha^{2}\left(-(t/\tau)^{\alpha}\right)$; see Eq. \ref{eqn:frm_Gt}. Nevertheless, we can use the standard technique of fitting $G(t)$ with a relaxation spectrum to infer $G^{*}(\omega)$.

\medskip

Figure \ref{fig:freqProp} shows $\epsilon^{*} = \epsilon^{\prime} - i \epsilon^{\prime\prime}$ and $G^{*} = G^{\prime} + i G^{\prime\prime}$ corresponding to the $P(t)$ and $G(t)$ depicted in figure \ref{fig:dynaProp}A. The real or storage component of these complex functions is denoted by a prime ($\epsilon^{\prime}$ and $G^{\prime}$) while the imaginary or loss component is indicated with a double prime ($\epsilon^{\prime\prime}$ and $G^{\prime\prime}$). Eq. \ref{eqn:epsilon_star} is used to obtain $\epsilon^{*}$, 
while the computer program pyReSpect is used to infer a relaxation spectrum from $G(t) = \sum_i g_i e^{-t/\tau_i}$.\cite{Shanbhag2019respect} The complex modulus shown in figure \ref{fig:freqProp}B is then obtained via $G^{*}(\omega) = \sum_{i} i \omega g_i \tau_i/(1 + i \omega \tau_i)$. Both $\epsilon^{*}$ and $G^{*}$ exhibit two distinct regimes of fast (high $\omega$) and slow (low $\omega$) relaxation modes.

\subsection{Generalized Memory Kernels}
\label{sec:generalized_memory_kernels}

Based on empirical observations in vitrimer systems, we proposed $K(t) \sim t^{-\alpha}$. However, most power-laws observed in real systems persist over a finite range of timescales, $\tau_{l} \leq t \leq \tau_{u}$, where $\tau_l$ and $\tau_u$ represent the lower and upper bounds. At shorter timescales $t < \tau_l$, the signal often saturates to a finite value instead of $K(0) \rightarrow \infty$, as implied by the power-law. Similarly, at longer timescales $t > \tau_u$, the memory kernel either saturates to a plateau for viscoelastic solids, such as gels and networks, or decays exponentially for viscoelastic liquids.

To generalize beyond power-law memory kernels, let us consider the Laplace transform (indicated by tilde) of the evolution equation for the $p$th mode, Eq. \ref{eqn:GLE_nmp_firm}:
\begin{equation}
\tilde{\mathbf{X}}_p(s) =  \tilde{\phi}_p(s) \mathbf{X}_p(0) + \tilde{G}_p(s) \tilde{\mathbf{F}}_p(s)
\label{eqn:laplace_evol}
\end{equation}
where $\tilde{\phi}_p(s) = \tilde{K}(s)/(s\tilde{K}(s) + \lambda_p)$ is the Laplace transform of the fundamental relaxation function, and $\tilde{G}_p(s) = 1/(s\tilde{K}(s) + \lambda_p)$ is the Laplace transform of the Green's function. By invoking linearity and the convolution theorem on Eq. \ref{eqn:laplace_evol}, an explicit time-domain Volterra equation emerges as:
\begin{equation}
\mathbf{X}_p(t) = \phi_p(t)\, \mathbf{X}_p(0) + \int_{0}^{t} G_p(t-t')\, \mathbf{F}_p(t') dt'.
\label{eqn:volterra}
\end{equation}
It turns out that the relaxation function $\phi_p(t) = \ave{\mathbf{X}_p(t)\cdot \mathbf{X}_p(0)}/\ave{X_p^2(0)}$. If a closed-form expression for $\phi_p(t)$ can be obtained via inverse Laplace transform, then Eqs. \ref{eqn:frm_Gt}, \ref{eqn:Ree_auto}, and  \ref{eqn:msd_n} can be generalized as
\begin{equation}
\begin{gathered}
G(t)/G_0 = \dfrac{1}{N} \sum_{p=1}^{N-1} \phi_{p}^{2}(t)\\
P(t) =  \dfrac{3 k_B T}{Nb^2} \sum_{p=1}^{N-1}  \dfrac{c_p^2}{\lambda_p} \phi_p(t)\\
\rho_n(t) - \rho_{\text{cm}}(t) = \sum_{p=1}^{N-1} \left(v_p^n \right)^2 \cdot \dfrac{6 k_B T}{\lambda_p} \left(1 - \phi_p(t) \right)\\
\rho_\text{cm}(t) = 6 k_B T v_0^2 \mathcal{L}^{-1}\left[1/\left(s^2 \tilde{K}(s) \right) \right]
\end{gathered}
\label{eqn:generalize}
\end{equation} 
where $\mathcal{L}^{-1}\left[\cdot\right]$ represents an inverse Laplace transform in the last equation. 

One approach for generalizing FIRM to model viscoelastic liquids is to use a tempered or truncated fractional memory kernel $K(t) \sim t^{-\alpha}\, e^{-t/\tau}$. However, this choice does not lead to a convenient analytical form for the relaxation function. A Prony series approximation $K(t) = \sum_{i=1}^{M} a_i e^{-b_i t}$ is another approach that is quite common for modeling GLEs. 
\begin{figure}
\begin{center}
\includegraphics[scale=0.6]{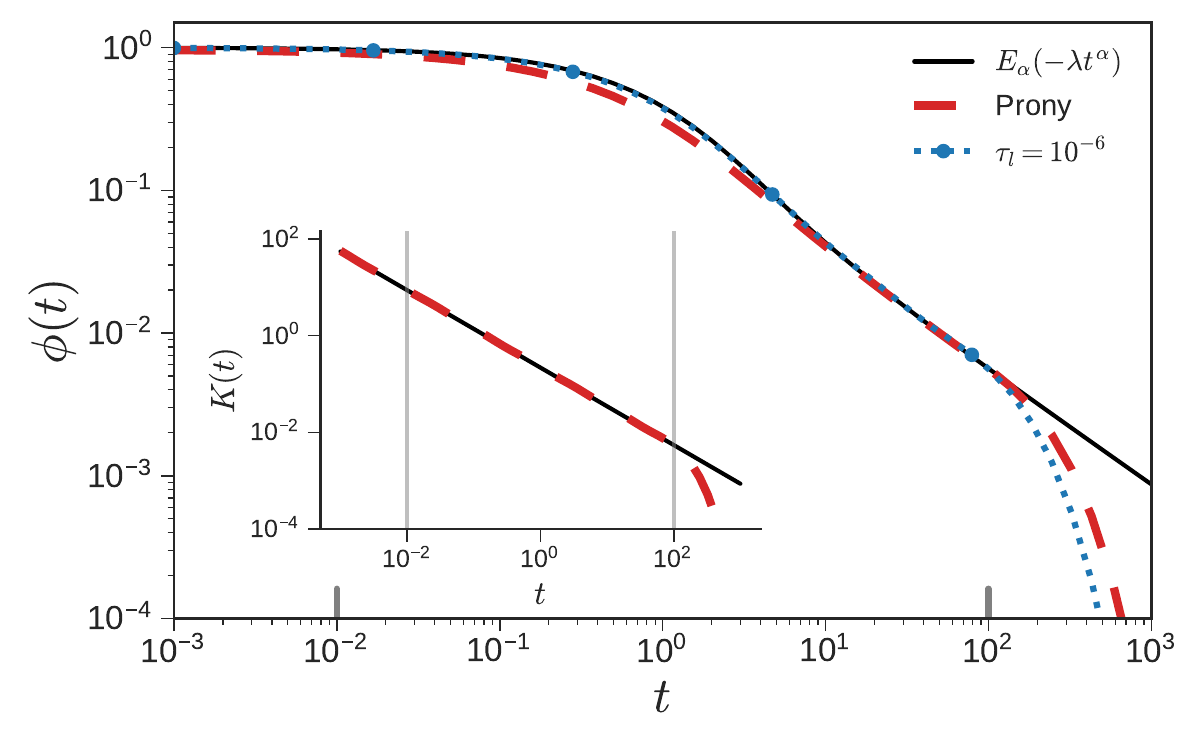}
\end{center}
\caption{Comparison of the relaxation function $\phi(t)$ of a Prony series approximation (dashed lines) over the range $t \in [10^{-2}, 10^{2}]$ (stems on horizontal axis) with the Mittag-Leffler function for $\lambda = 1$ and $\alpha = 0.8$. The inset compares the memory kernel $t^{-\alpha}/\Gamma(1-\alpha)$ with the Prony series approximation. The dotted line with circles shows the relaxation function for a Prony approximation over the range $t \in [10^{-6}, 10^{2}]$.\label{fig:prony}}
\end{figure}
An example is shown in the inset to figure \ref{fig:prony} for $\alpha = 0.8$. The Prony approximation for the power-law memory kernel between $\tau_l = 10^{-2}$ and $\tau_u = 10^2$ s is obtained using the program pyReSpect.\cite{Shanbhag2019respect} A parsimonious fit, shown in the inset, is obtained with $M = 10$ exponential modes.

The Laplace transform of such a kernel is a rational function,
\begin{equation}
\tilde{K}(s) = \sum_{i=1}^{M} \dfrac{a_i}{s + b_i} = \dfrac{P_{M-1}(s)}{Q_M(s)}
\label{eqn:Ks_prony}
\end{equation}
where $P_{M-1}(s)$ and $Q_M(s)$ polynomials of degree $M-1$ and $M$, respectively. Thus, the Laplace transform of the relaxation function is also a rational function which can be decomposed into partial fractions
\begin{equation}
\tilde{\phi}_p(s) = \dfrac{P_{M-1}(s)}{sP_{M-1}(s) + \lambda_p Q_{M}(s)} = \sum_{i=1}^{M} \dfrac{r_i}{s + p_i}
\end{equation}
where $r_i$ and $p_i$ are the residues and poles, respectively. This implies that the relaxation function is also a Prony series,
\begin{equation}
\phi_p(t) = \sum_{i=1}^{M} r_i e^{-p_i t}.
\label{eqn:phi_p_Prony}
\end{equation}

Determination of the relaxation function requires us to specify $\lambda_p$ in addition to the coefficients $\{a_i, b_i\}$. We set $\lambda = 1$ and used the scipy signal processing library to resolve $\tilde{\phi}_p(s)$ into partial fractions.\cite{Virtanen2020, Shanbhag2023_psi, Shanbhag2026} Figure \ref{fig:prony} compares the relaxation functions corresponding to the power-law, $K(t) = t^{-\alpha}/\Gamma(1-\alpha)$, and the Prony approximation, $K(t) = \sum_i a_i e^{-b_i t}$, kernels. The relaxation function corresponding to the power-law kernel is the Mittag-Leffler function, $E_{\alpha}(-\lambda t^{\alpha})$ which is shown by the solid line. The $\phi(t)$ computed from $\{r_i, p_i\}$ via Eq. \ref{eqn:phi_p_Prony} is shown by the dashed line. The two relaxation functions show good agreement, especially for $t \in [\tau_l, \tau_u]$.

The key difference between the two relaxation functions is that the $\phi(t)$ corresponding to the Prony series shows an exponential decay for $t \gtrsim \tau_u$. The agreement between the Mittag-Leffler function and the Prony approximation for $t \lesssim 1$ can be improved by decreasing $\tau_l$. Only two additional modes ($M=12$) are required to approximate $K(t)$ for $10^{-6} \leq t \leq 10^{2}$. The dotted line with circles in figure \ref{fig:prony} shows the resulting improvement in approximating the Mittag-Leffler function over this range.

\subsection{Physical Significance of Fractional Exponent}

FIRM provides a rigorous framework for describing the extended power-law relaxations observed in vitrimer systems. Within this framework, the characteristic power-law scaling exponent emerges naturally as the parameter $\alpha$. We propose that $\alpha$ offers additional insight into the microscopic processes that govern the slow relaxation regime in vitrimers. In the following paragraphs, we examine possible physical interpretations of this parameter.

One potential interpretation is that $\alpha$ reflects the kinetics for local cross-link exchange in the absence of cross-linker or polymer diffusion limitations. This hypothesis is motivated by our recent observations in imine-based PS vitrimers prepared with excess diamine cross-linkers whose $pK_{a}$ values – a measure of amine nucleophilicity – ranged from 3.0 to 10.5. During creep measurements, all PS vitrimers exhibited extended power-law behavior in the creep compliance. The measured power-law scaling exponents varied linearly with the diamine cross-linker $pK_{a}$ but showed no correlation with the cross-linker diffusion coefficient.\cite{Barzycki2026} Based on this result, we speculate that $\alpha$ potentially provides a rheological measure of how vitrimer chemistry influences the kinetic energy barriers that dictate local cross-link exchange.

Conversely, $\alpha$ also potentially describes the sub-diffusive mobility of reactive sites within polymer networks. Neutron spin-echo studies of polydimethylsiloxane networks found that the MSD of the permanent cross-links exhibited a Rouse-like power-law ($\alpha = 0.5$) at short and intermediate times, which plateaued at long times.\cite{Oeser1988, Richter1989} In transient networks, cross-links are spatially unrestricted (e.g. figure \ref{fig:dynaProp}B) and the MSD does not plateau at long times. In real networks, the sub-diffusion exponent may not be constant and can take on different values over different timescales. Molecular dynamics simulations can be used to probe the MSD,\cite{Perego2022} which can then be incorporated into FIRM via generalized memory kernels.

The fractional exponent $\alpha$ is also important because it controls the effective or renormalized lifetime of bonds between reactive sites in dissociative networks.\cite{Rubinstein1998, Shanbhag2023} Subdiffusive processes ($\alpha < 1$) lead to compact random walks: bonds between stickers break and reform several times before finding new partners. This geometric process is neglected in this work. We assumed that the cross-link lifetime $\tau_x$ follows an Arrhenius relation with energy barrier $E_a$, and an attempt frequency set by the segmental relaxation time. However, the renormalized bond lifetime model shows that $\alpha$ modifies the effective attempt frequency.\cite{Ricarte2023} We speculate that inclusion of these neglected geometric processes, and their dependence on temperature, may help to partially resolve the temperature-dependent $E_a$ required to reconcile experimental data in figure \ref{fig:PS_Gt_fit}.

\section{Summary and Conclusions}

Motivated by the observation that unentangled vitrimers and other transient polymer networks routinely exhibit stress relaxation power-laws with exponents that differ from the $t^{-1/2}$ scaling of the sticky Rouse model,\cite{Ricarte2021} we developed the FIRM, combining two previously separate generalizations of the Rouse model: IHR, which allows beads to carry different friction coefficients to represent sticky cross-links,\cite{Ricarte2021} and the fractional Rouse model (FRM), in which beads are driven by fractional Gaussian noise rather than ordinary Brownian kicks to capture the subdiffusive dynamics of a chain embedded in a viscoelastic matrix.\cite{Sharma2010} Thus, FIRM provides a general framework that interpolates between FRM for homogeneous chains and IHR in the Markovian limit ($\alpha = 1$). We also introduced the FSRM, a two-population analytical approximation to FIRM, and found that it reproduces FIRM faithfully when the number of sticky beads $N_x$ and the friction contrast $\delta = \zeta_x/\zeta$ are both large.

The relaxation modulus predicted by FIRM retains the mathematical structure of the SRM and IHR, but replaces exponential relaxation with the Mittag-Leffler function. Rather than a single $t^{-1/2}$ transition regime, FIRM predicts two distinct power-laws flanking the Rouse relaxation time, $G(t) \sim t^{-\alpha/2}$ at intermediate times and $G(t) \sim t^{-2\alpha}$ at long times. For cross-linked chains, an analogous pair of power-laws appears on either side of the cross-link lifetime $\tau_x$. This behavior is insensitive to whether the sticky beads are distributed uniformly or randomly along the backbone. 
 
Fitting FIRM to stress relaxation measurements on an imine-based polystyrene vitrimer (PS-v-8-1.2) across a range of temperatures shows that the subdiffusive exponent $\alpha$ is essential for capturing the shape of the extended terminal relaxation. The best-fitting values of $\alpha$ were confined to the range ($0.85 \pm 0.03$) across all temperatures probed. $E_a$ may be treated as a constant or increasing with temperature, representative of Arrhenius or non-Arrhenius behavior, respectively. We speculate that non-Arrhenius behavior may potentially result from geometric processes that modify the attempt frequency in the Arrhenius relation, consistent with prior evidence that cross-link exchange may not be purely reaction-limited.\cite{Barzycki2025, Barzycki2026}
 
Because FIRM is a microscopic model rather than a purely phenomenological fitting function like fractional viscoelastic models, it is not restricted to linear shear rheology. We derived closed-form expressions for the autocorrelation of the end-to-end vector, the mean-squared displacement of individual beads and the chain center-of-mass, and the associated dielectric response, all expressed in terms of the same generalized eigenmodes that determine $G(t)$. These relations point toward a self-consistent way of combining rheological, dielectric, and scattering measurements to constrain FIRM parameters, which may help disentangle segmental and cross-link exchange contributions to the relaxation spectrum of a given system. We also outlined how the idealized power-law memory kernel $K(t) \sim t^{-\alpha}$ can be generalized to tempered or Prony-series kernels that saturate or decay outside a finite range of timescales, providing a route to extend FIRM to viscoelastic solids and liquids whose subdiffusive character is not scale-invariant over all times.

\section*{Acknowledgements}
We thank Daniel Barzycki for assistance with organizing the experimental imine-based polystyrene vitrimer stress relaxation data. Sachin Shanbhag acknowledges support from the Department of Scientific Computing at Florida State University. Ralm Ricarte was primarily supported by the National Science Foundation (DMR-2144007).

\section*{Author Declarations}

\subsection*{Conflict of Interest}

The authors have no conflicts to disclose.

\subsection*{Author Contributions}

\noindent
\textbf{Sachin Shanbhag}: Conceptualization (lead), Formal Analysis (lead), Investigation (equal), Project Administration (lead), Writing (equal). \textbf{Ralm G. Ricarte}: Formal Analysis (supporting), Investigation (equal), Writing (equal).

\section*{Data Availability Statement}
The data that support the findings of this study are available from the corresponding author upon reasonable request.

\appendix

\section{Comparison with a Fractional Maxwell Model}
\label{sec:fmm}

The transition between the two power-laws observed in figure \ref{fig:frm} of the manuscript is reminiscent of the fractional Maxwell model (FMM), which is a phenomenological consisting of two springpots.\cite{Friedrich1991, Song2023} It is characterized by four non-negative parameters $(c_a, a)$ and $(c_b, b)$ where the two sets of parameters correspond to the two springpots. $a$ and $b$ are constrained by $0 \leq b \leq a \leq 1$. 

\begin{figure}
\begin{center}
\includegraphics[scale=0.6]{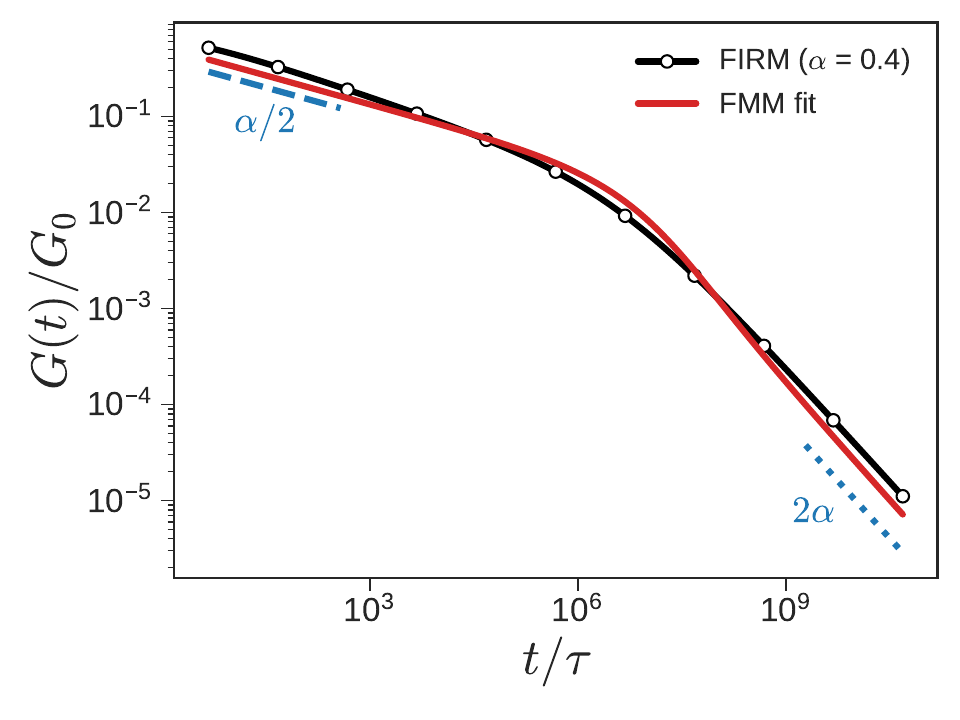}
\end{center}
\caption{A FMM (red line) is used to fit the $G(t)$ computed using the FRM (black line with circle markers) with $N=20$ and $\alpha=0.4$. The parameters of the FMM fit are $c_a/G_0 = 25.58$ s${}^{0.8}$, $c_b/G_0 = 0.14$ s${}^{0.2}$, $a = 0.8$ and $b = 0.2$.\label{fig:fmm_fit}}
\end{figure}

The FMM exhibits a transition from $G(t) \sim t^{-b}$ at short timescales to $G(t) \sim t^{-a}$ at long timescales given by the expression
\begin{equation}
G_\text{FMM}(t) = c_b t^{-b} E_{a-b, 1-b}\left(-\dfrac{c_b}{c_a} t^{a-b}\right),
\label{eqn:fmm_Gt}
\end{equation}
where $E_{\alpha, \beta}(\cdot)$ is the two parameter Mittag-Leffler function.\cite{Friedrich1991} The quasi-properties $c_a$ and $c_b$ have units of Pa $\cdot$ s${}^{a}$ and Pa $\cdot$ s${}^{b}$, respectively. Can the response of the FRM be concisely described by the FMM? The FRM shows a transition from  $b=\alpha/2$ to $a=2\alpha$. However the condition $a \leq 1$ in the FMM implies that $\alpha < 1/2$ in the FRM. This implies that the viscosity of the FMM always diverges.

Nevertheless, it is interesting to test the agreement between the FRM and FMM for $\alpha < 0.5$. In figure \ref{fig:fmm_fit}, we computed $G(t)$ of a FRM with $N = 20$ and $\alpha = 0.4$. It shows the transition from $G(t) \sim t^{-0.2}$ to $G(t) \sim t^{-0.8}$ as expected. We fit this $G(t)$ to a FMM by setting $a = 2\alpha = 0.8$ and $b = \alpha/2 = 0.2$. Using nonlinear least squares regression, we determine the best fitting values for the remaining two parameters as $c_a/G_0 = 13.0$ s${}^{0.8}$ and $c_b/G_0 = 0.11$ s${}^{0.2}$. As expected, the shape of the FMM roughly matches the shape of $G(t)$ determined by FRM for $\alpha < 0.5$. Despite the same terminal slopes, the two curves in figure \ref{fig:fmm_fit} do not overlap. Thus, the behavior of the FRM cannot be quantitatively captured by a FMM.

\section{Fitting Experiments with Constant $E_a$}
\label{sec:fit_Ea35}

\begin{figure}
\begin{center}
\includegraphics[scale=0.6]{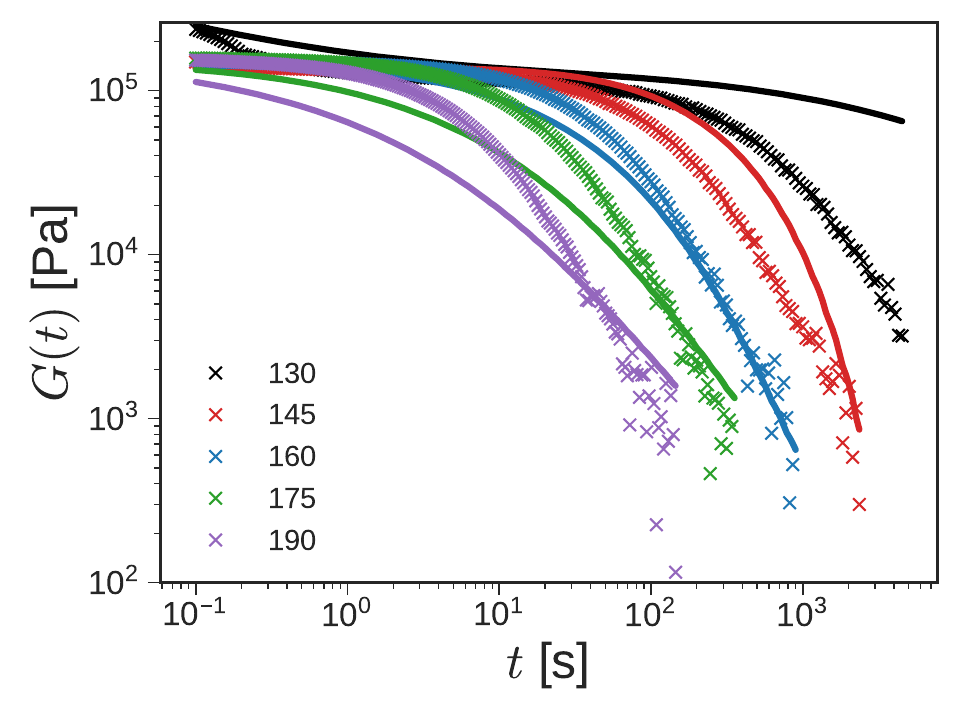}
\end{center}
\caption{Fits of the FIRM to PS-v-8-1.2 data (+) at different temperatures with $\alpha$ as a free parameter and constant $E_a$ = 35 kJ/mol.\label{fig:EaConstant}}
\end{figure}

Figure \ref{fig:EaConstant} shows fits of the FIRM with adjustable $\alpha$ and constant $E_a$. The chosen value of $E_a$ = 35 kJ/mol was approximately the average value of $E_a$ when it was also allowed to be a free parameter (figure \ref{fig:PS_Gt_fit}B). The quality of the fits is poor; it cannot be improved much by varying the fixed value of $E_a$. This offers further evidence that the underlying mechanism of cross-link exchange is non-Arrhenius. Furthermore, we observe no discernible pattern in the fitted values of $\alpha$ with temperature. The regressed values of $\alpha$ were 0.5 (130 ${}^\circ$C), 1.0 (145 ${}^\circ$C), 0.78 (160 ${}^\circ$C), 0.62 (175 ${}^\circ$C), and 0.55 (190 ${}^\circ$C). Thus, both $\alpha$ and $E_a$ need to be adjustable to properly describe the data.

\bibliography{firm.bib}

@article{Stern1946,
  author =        {M. D. Stern and A. V. Tobolsky},
  journal =       {Rubber Chem. Technol.},
  number =        {4},
  pages =         {1178--1192},
  title =         {Stress-Time-Temperature Relations in Polysulfide
                   Rubbers},
  volume =        {19},
  year =          {1946},
  doi =           {10.5254/1.3543255},
  issn =          {0035-9475},
}

@article{Green1946,
  author =        {Green, M. S. and Tobolsky, A. V.},
  journal =       {J. Chem. Phys.},
  month =         feb,
  number =        {2},
  pages =         {80--92},
  title =         {{A new approach to the theory of relaxing polymeric
                   media}},
  volume =        {14},
  year =          {1946},
  doi =           {10.1063/1.1724109},
  issn =          {0021-9606},
}

@book{Tanaka_2011,
  address =       {Cambridge, UK},
  author =        {Tanaka, Fumihiko},
  publisher =     {Cambridge University Press},
  title =         {Polymer Physics: {Applications} to Molecular
                   Association and Thermoreversible Gelation},
  year =          {2011},
  doi =           {10.1017/CBO9780511975691},
}

@article{Webber2022,
  author =        {Webber, Matthew J. and Tibbitt, Mark W.},
  journal =       {Nature Reviews Materials},
  number =        {7},
  pages =         {541--556},
  title =         {Dynamic and reconfigurable materials from reversible
                   network interactions},
  volume =        {7},
  year =          {2022},
  doi =           {10.1038/s41578-021-00412-x},
  issn =          {2058-8437},
}

@article{Ricarte2026,
  author =        {Ricarte, Ralm G.},
  journal =       {Macromolecules},
  number =        {12},
  pages =         {6593-6619},
  title =         {Network Topology and Linear Viscoelasticity of
                   Vitrimers -- {A} Polymer Physics Perspective},
  volume =        {59},
  year =          {2026},
  doi =           {10.1021/acs.macromol.6c00582},
}

@article{Montarnal2011,
  author =        {Montarnal, Damien and Capelot, Mathieu and
                   Tournilhac, Fran{\c{c}}ois and Leibler, Ludwik},
  journal =       {Science},
  number =        {6058},
  pages =         {965--968},
  title =         {Silica-like malleable materials from permanent
                   organic networks},
  volume =        {334},
  year =          {2011},
  doi =           {10.1126/science.1212648},
}

@article{Winne2019,
  author =        {Winne, Johan M. and Leibler, Ludwik and
                   Du Prez, Filip E.},
  journal =       {Polym. Chem.},
  pages =         {6091--6108},
  publisher =     {The Royal Society of Chemistry},
  title =         {Dynamic covalent chemistry in polymer networks: a
                   mechanistic perspective},
  volume =        {10},
  year =          {2019},
  doi =           {10.1039/C9PY01260E},
}

@article{McBride2019,
  author =        {McBride, Matthew K. and Worrell, Brady T. and
                   Brown, Tobin and Cox, Lewis M. and Sowan, Nancy and
                   Wang, Chen and Podgorski, Maciej and
                   Martinez, Alina M. and Bowman, Christopher N.},
  journal =       {Ann. Rev. Chem. Biomol. Engg.},
  number =        {1},
  pages =         {175--198},
  title =         {Enabling applications of covalent adaptable networks},
  volume =        {10},
  year =          {2019},
  doi =           {10.1146/annurev-chembioeng-060718-030217},
}

@article{VanZee2020,
  author =        {{Van Zee}, Nathan J. and Nicola{\"{y}}, Renaud},
  journal =       {Prog. Polym. Sci.},
  pages =         {101233},
  publisher =     {Elsevier Ltd},
  title =         {{Vitrimers: {P}ermanently crosslinked polymers with
                   dynamic network topology}},
  volume =        {104},
  year =          {2020},
  doi =           {10.1016/j.progpolymsci.2020.101233},
  issn =          {00796700},
}

@article{Hayashi2025,
  author =        {Mikihiro Hayashi and Ralm G. Ricarte},
  journal =       {Prog. Polym. Sci.},
  pages =         {102026},
  title =         {Towards the next development of vitrimers: {Recent}
                   key topics for the practical application and
                   understanding of the fundamental physics},
  volume =        {170},
  year =          {2025},
  doi =           {https://doi.org/10.1016/j.progpolymsci.2025.102026},
  issn =          {0079-6700},
}

@article{Williams1955,
  author =        {Williams, Malcolm L. and Landel, Robert F. and
                   Ferry, John D.},
  journal =       {J. Am. Chem. Soc.},
  number =        {14},
  pages =         {3701--3707},
  title =         {The temperature dependence of relaxation mechanisms
                   in amorphous polymers and other glass-forming
                   liquids},
  volume =        {77},
  year =          {1955},
  doi =           {10.1021/ja01619a008},
}

@article{Meng2022,
  author =        {Meng, Fanlong and Saed, Mohand O. and
                   Terentjev, Eugene M.},
  journal =       {Nat. Commun.},
  number =        {1},
  pages =         {5753},
  title =         {Rheology of vitrimers},
  volume =        {13},
  year =          {2022},
  doi =           {10.1038/s41467-022-33321-w},
}

@article{Lin2025,
  author =        {Lin, Tsai-Wei and Mei, Baicheng and Dutta, Sarit and
                   Schweizer, Kenneth S. and Sing, Charles E.},
  journal =       {Macromolecules},
  number =        {3},
  pages =         {1481-1497},
  title =         {Molecular Dynamics Simulation and Theoretical
                   Analysis of Structural Relaxation, Bond Exchange
                   Dynamics, and Glass Transition in Vitrimers},
  volume =        {58},
  year =          {2025},
  doi =           {10.1021/acs.macromol.4c02659},
}

@article{Wang2026,
  author =        {Wang, Hao and Jiang, Nuofei and Zhang, Rongchun and
                   Nicolay, Renaud and Fustin, Charles-Andr{\'e} and
                   Van Ruymbeke, Evelyne},
  journal =       {Macromolecules},
  number =        {5},
  pages =         {2818-2834},
  title =         {Linear Viscoelasticity of Model Vitrimers: {Decoding}
                   the Interplay between Bond Exchange and Segmental
                   Dynamics},
  volume =        {59},
  year =          {2026},
  doi =           {10.1021/acs.macromol.5c03092},
}

@article{Quinteros-Sedano2026,
  author =        {Quinteros-Sedano, Alvaro and van Ruymbeke, Evelyne},
  journal =       {Macromolecules},
  number =        {3},
  pages =         {1781-1788},
  title =         {Tuning the Viscoelastic Properties of Dioxazaborocane
                   Vitrimers via Chemical Design of the Functional
                   Groups},
  volume =        {59},
  year =          {2026},
  doi =           {10.1021/acs.macromol.5c03243},
}

@article{Ricarte2023,
  author =        {Ricarte, Ralm G. and Shanbhag, Sachin and
                   Ezzeddine, Dana and Barzycki, Daniel and Fay, Kevin},
  journal =       {Macromolecules},
  number =        {17},
  pages =         {6806--6817},
  title =         {Time–temperature superposition of polybutadiene
                   vitrimers},
  volume =        {56},
  year =          {2023},
  doi =           {10.1021/acs.macromol.3c00883},
}

@article{Barzycki2025,
  author =        {Barzycki, Daniel C. and Ezzeddine, Dana and
                   Shanbhag, Sachin and Ricarte, Ralm G.},
  journal =       {Macromolecules},
  number =        {8},
  pages =         {3949-3963},
  title =         {Linear Viscoelasticity of Polystyrene Vitrimers:
                   {Segmental} Motions and the Slow Arrhenius Process},
  volume =        {58},
  year =          {2025},
  doi =           {10.1021/acs.macromol.4c03161},
}

@article{Barzycki2026,
  author =        {Barzycki, Daniel C. and Ezzeddine, Dana and
                   Shanbhag, Sachin and Ricarte, Ralm G.},
  journal =       {ACS Applied Polymer Materials},
  number =        {4},
  pages =         {2964-2974},
  title =         {Interplay between Cross-Linker Nucleophilicity and
                   Diffusion in Polystyrene Vitrimer Dynamics},
  volume =        {8},
  year =          {2026},
  doi =           {10.1021/acsapm.5c04468},
}

@article{Quinteros-Sedano2025,
  author =        {Quinteros-Sedano, Alvaro and Le Besnerais, Brieuc and
                   Van Zee, Nathan J. and Nicolaÿ, Renaud},
  journal =       {Chem. Mater.},
  number =        {5},
  pages =         {2058-2070},
  title =         {Exploiting Dioxazaborocane Chemistry for Preparing
                   Elastomeric Vitrimers with Enhanced Processability
                   and Mechanical Properties},
  volume =        {37},
  year =          {2025},
  doi =           {10.1021/acs.chemmater.5c00154},
}

@article{Martins2023,
  author =        {Martins, Murillo L. and Zhao, Xiao and
                   Demchuk, Zoriana and Luo, Jiancheng and
                   Carden, Gregory Peyton and Toleutay, Gaukhar and
                   Sokolov, Alexei P.},
  journal =       {Macromolecules},
  number =        {21},
  pages =         {8688-8696},
  title =         {Viscoelasticity of Polymers with Dynamic Covalent
                   Bonds: {Concepts} and Misconceptions},
  volume =        {56},
  year =          {2023},
  doi =           {10.1021/acs.macromol.3c01545},
}

@article{rouse53,
  author =        {P. R. Rouse},
  journal =       {J. Chem. Phys.},
  pages =         {1272--1280},
  title =         {A theory of the linear viscoelastic properties of
                   dilute solutions of coiling polymers},
  volume =        {21},
  year =          {1953},
}

@article{Baxandall1989,
  author =        {Baxandall, L. G.},
  journal =       {Macromolecules},
  number =        {4},
  pages =         {1982--1988},
  title =         {Dynamics of reversibly crosslinked chains},
  volume =        {22},
  year =          {1989},
  doi =           {10.1021/ma00194a076},
  issn =          {15205835},
}

@article{Leibler1991,
  author =        {Leibler, Ludwik and Rubinstein, Michael and
                   Colby, Ralph H.},
  journal =       {Macromolecules},
  number =        {16},
  pages =         {4701--4707},
  title =         {Dynamics of reversible networks},
  volume =        {24},
  year =          {1991},
  doi =           {10.1021/ma00016a034},
}

@article{Rubinstein1998,
  author =        {Rubinstein, Michael and Semenov, Alexander N.},
  journal =       {Macromolecules},
  number =        {4},
  pages =         {1386--1397},
  title =         {Thermoreversible gelation in solutions of associating
                   polymers. {2. Linear} dynamics},
  volume =        {31},
  year =          {1998},
  doi =           {10.1021/ma970617+},
}

@article{Chen2013,
  author =        {Chen,Quan and Tudryn,Gregory J. and Colby,Ralph H.},
  journal =       {J. Rheol.},
  number =        {5},
  pages =         {1441--1462},
  title =         {Ionomer dynamics and the sticky {Rouse} model},
  volume =        {57},
  year =          {2013},
  doi =           {10.1122/1.4818868},
}

@article{Chen2016,
  author =        {Chen,Quan and Zhang,Zhijie and Colby,Ralph H.},
  journal =       {J. Rheol.},
  number =        {6},
  pages =         {1031--1040},
  title =         {Viscoelasticity of entangled random polystyrene
                   ionomers},
  volume =        {60},
  year =          {2016},
  doi =           {10.1122/1.4955432},
}

@article{Ricarte2021,
  author =        {Ricarte, Ralm G. and Shanbhag, Sachin},
  journal =       {Macromolecules},
  number =        {7},
  pages =         {3304--3320},
  title =         {Unentangled vitrimer melts: {I}nterplay between chain
                   relaxation and cross-link exchange controls linear
                   rheology},
  volume =        {54},
  year =          {2021},
  doi =           {10.1021/acs.macromol.0c02530},
}

@article{Jiang2021,
  author =        {Jiang, Nuofei and Zhang, Hongdong and Yang, Yuliang and
                   Tang, Ping},
  journal =       {J. Rheol.},
  number =        {4},
  pages =         {527-547},
  title =         {Molecular dynamics simulation of associative
                   polymers: {Understanding} linear viscoelasticity from
                   the sticky {Rouse} model},
  volume =        {65},
  year =          {2021},
  doi =           {10.1122/8.0000218},
}

@article{Shao2022,
  author =        {Shao, Jingyu and Jiang, Nuofei and Zhang, Hongdong and
                   Yang, Yuliang and Tang, Ping},
  journal =       {Macromolecules},
  number =        {2},
  pages =         {535-549},
  title =         {Sticky {Rouse} Model and Molecular Dynamics
                   Simulation for Dual Polymer Networks},
  volume =        {55},
  year =          {2022},
  doi =           {10.1021/acs.macromol.1c02059},
}

@article{Cui2023,
  author =        {Cui, Xiang and Jiang, Nuofei and Shao, Jingyu and
                   Zhang, Hongdong and Yang, Yuliang and Tang, Ping},
  journal =       {Macromolecules},
  number =        {3},
  pages =         {772--784},
  title =         {Linear and nonlinear viscoelasticities of
                   dissociative and associative covalent adaptable
                   networks: {Discrepancies} and limits},
  volume =        {56},
  year =          {2023},
  doi =           {10.1021/acs.macromol.2c02122},
}

@article{Cui2025,
  author =        {Cui, Xiang and Luo, Yulin and Yang, Yuliang and
                   Tang, Ping},
  journal =       {Macromolecules},
  number =        {4},
  pages =         {1898-1911},
  title =         {Decoupling Multiple Relaxation Modes: {Composition}
                   and Distribution of Terminal Relaxation Time in
                   Associative Polymers},
  volume =        {58},
  year =          {2025},
  doi =           {10.1021/acs.macromol.4c02349},
}

@article{Zhang2026,
  author =        {Zhang, Lu and Yang, Yuliang and Zhang, Hongdong and
                   Tang, Ping},
  journal =       {Macromolecules},
  number =        {12},
  pages =         {7183-7198},
  title =         {Topological Decoupling of Association Kinetics and
                   Viscoelasticity in Associative Star Polymers: A
                   Sticky {Rouse} Model and Simulation Study},
  volume =        {59},
  year =          {2026},
  doi =           {10.1021/acs.macromol.6c00423},
}

@book{doipd,
  address =       {Oxford},
  author =        {M. Doi and S. F. Edwards},
  publisher =     {Clarendon Press},
  title =         {The theory of polymer dynamics},
  year =          {1986},
}

@article{Langevin1908,
  author =        {Langevin, Paul and others},
  journal =       {C. R. Acad. Sci. Paris},
  pages =         {530-533},
  title =         {Sur la th{\'e}orie du mouvement brownien},
  volume =        {146},
  year =          {1908},
}

@article{Mori1965a,
  author =        {Mori, Hazime},
  journal =       {Prog. Theor. Phys.},
  month =         {03},
  number =        {3},
  pages =         {423-455},
  title =         {Transport, Collective Motion, and Brownian Motion*},
  volume =        {33},
  year =          {1965},
  doi =           {10.1143/PTP.33.423},
  issn =          {0033-068X},
}

@article{Zwanzig1973,
  author =        {Zwanzig, Robert},
  journal =       {J. Stat. Phys.},
  number =        {3},
  pages =         {215--220},
  title =         {Nonlinear generalized {Langevin} equations},
  volume =        {9},
  year =          {1973},
  doi =           {10.1007/BF01008729},
  issn =          {1572-9613},
}

@article{Lee2000,
  author =        {Lee, M. Howard},
  journal =       {Phys. Rev. E},
  pages =         {1769--1772},
  publisher =     {American Physical Society},
  title =         {Generalized {Langevin} equation and recurrence
                   relations},
  volume =        {62},
  year =          {2000},
  doi =           {10.1103/PhysRevE.62.1769},
}

@article{McKinley2009,
  author =        {McKinley, Scott A. and Yao, Lingxing and
                   Forest, M. Gregory},
  journal =       {J. Rheol.},
  month =         nov,
  number =        {6},
  pages =         {1487--1506},
  title =         {Transient anomalous diffusion of tracer particles in
                   soft matter},
  volume =        {53},
  year =          {2009},
  doi =           {10.1122/1.3238546},
  issn =          {0148-6055},
}

@article{McKinley2018,
  author =        {McKinley, Scott A. and Nguyen, Hung D.},
  journal =       {SIAM J. Math. Anal.},
  number =        {5},
  pages =         {5119--5160},
  title =         {Anomalous diffusion and the generalized {Langevin}
                   equation},
  volume =        {50},
  year =          {2018},
  doi =           {10.1137/17M115517X},
}

@article{Mason1995,
  author =        {Mason, T. G. and Weitz, D. A.},
  journal =       {Phys. Rev. Lett.},
  pages =         {1250--1253},
  publisher =     {American Physical Society},
  title =         {Optical Measurements of Frequency-Dependent Linear
                   Viscoelastic Moduli of Complex Fluids},
  volume =        {74},
  year =          {1995},
  doi =           {10.1103/PhysRevLett.74.1250},
  url =           {https://link.aps.org/doi/10.1103/PhysRevLett.74.1250},
}

@article{Mason1997,
  author =        {Mason, T. G. and Ganesan, K. and van Zanten, J. H. and
                   Wirtz, D. and Kuo, S. C.},
  journal =       {Phys. Rev. Lett.},
  pages =         {3282--3285},
  publisher =     {American Physical Society},
  title =         {Particle Tracking Microrheology of Complex Fluids},
  volume =        {79},
  year =          {1997},
  doi =           {10.1103/PhysRevLett.79.3282},
}

@article{Fricks2009,
  author =        {Fricks, John and Yao, Lingxing and Elston, Timothy C. and
                   Forest, M. Gregory},
  journal =       {SIAM J. Appl. Math.},
  number =        {5},
  pages =         {1277--1308},
  title =         {Time-domain methods for diffusive transport in soft
                   matter},
  volume =        {69},
  year =          {2009},
  doi =           {10.1137/070695186},
}

@article{Cordoba2012,
  author =        {C\'{o}rdoba, Andrés and Indei, Tsutomu and
                   Schieber, Jay D.},
  journal =       {J. Rheol.},
  month =         jan,
  number =        {1},
  pages =         {185--212},
  title =         {Elimination of inertia from a generalized {Langevin}
                   equation: applications to microbead rheology modeling
                   and data analysis},
  volume =        {56},
  year =          {2012},
  doi =           {10.1122/1.3675625},
  issn =          {0148-6055},
}

@article{Mandelbrot1968,
  author =        {Mandelbrot, Benoit B. and Van Ness, John W.},
  number =        {4},
  journal = {SIAM Rev.},
  pages =         {422--437},
  year  = {1968},
  title =         {Fractional Brownian Motions, Fractional Noises and
                   Applications},
  volume =        {10},
  doi =           {10.1137/1010093},
}

@article{Barkai2000,
  author =        {Barkai, E. and Metzler, R. and Klafter, J.},
  journal =       {Phys. Rev. E},
  month =         jan,
  pages =         {132--138},
  publisher =     {American Physical Society},
  title =         {From continuous time random walks to the fractional
                   {Fokker-Planck} equation},
  volume =        {61},
  year =          {2000},
  doi =           {10.1103/PhysRevE.61.132},
}

@article{Shanbhag2023,
  author =        {Shanbhag, Sachin and Ricarte, Ralm G.},
  journal =       {Macromol. Theory Simul.},
  number =        {4},
  pages =         {2300002},
  title =         {On the effective lifetime of reversible bonds in
                   transient networks},
  volume =        {32},
  year =          {2023},
  doi =           {10.1002/mats.202300002},
}

@article{Weber1993,
  author =        {Weber, Hans Werner and Kimmich, Rainer},
  journal =       {Macromolecules},
  number =        {10},
  pages =         {2597--2606},
  title =         {Anomalous segment diffusion in polymers and {NMR}
                   relaxation spectroscopy},
  volume =        {26},
  year =          {1993},
  doi =           {10.1021/ma00062a031},
}

@article{Wong2004,
  author =        {Wong, I. Y. and Gardel, M. L. and Reichman, D. R. and
                   Weeks, Eric R. and Valentine, M. T. and Bausch, A. R. and
                   Weitz, D. A.},
  journal =       {Phys. Rev. Lett.},
  month =         apr,
  pages =         {178101},
  publisher =     {American Physical Society},
  title =         {Anomalous diffusion probes microstructure dynamics of
                   entangled {F}-actin networks},
  volume =        {92},
  year =          {2004},
  doi =           {10.1103/PhysRevLett.92.178101},
}

@article{Tang2015a,
  author =        {Tang, Shengchang and Wang, Muzhou and
                   Olsen, Bradley D.},
  journal =       {J. Am. Chem. Soc.},
  month =         mar,
  number =        {11},
  pages =         {3946--3957},
  title =         {Anomalous self-diffusion and sticky {Rouse} dynamics
                   in associative protein hydrogels},
  volume =        {137},
  year =          {2015},
  doi =           {10.1021/jacs.5b00722},
  issn =          {15205126},
}

@article{Panja2010,
  author =        {Panja, Debabrata},
  journal =       {J. Stat. Mech: Theory Exp.},
  number =        {02},
  pages =         {L02001},
  title =         {Generalized {Langevin} equation formulation for
                   anomalous polymer dynamics},
  volume =        {2010},
  year =          {2010},
  doi =           {10.1088/1742-5468/2010/02/L02001},
}

@article{Tian2022,
  author =        {Tian, Xiaofei and Xu, Xiaolei and Chen, Ye and
                   Chen, Jizhong and Xu, Wen-Sheng},
  journal =       {J. Chem. Phys.},
  month =         {12},
  number =        {22},
  pages =         {224901},
  title =         {Explicit analytical form for memory kernel in the
                   generalized {Langevin equation} for end-to-end vector
                   of {Rouse} chains},
  volume =        {157},
  year =          {2022},
  doi =           {10.1063/5.0124925},
  issn =          {0021-9606},
}

@article{Kou2004,
  author =        {Kou, S. C. and Xie, X. Sunney},
  journal =       {Phys. Rev. Lett.},
  pages =         {180603},
  title =         {Generalized {Langevin} equation with fractional
                   {Gaussian} noise: {Subdiffusion} within a single
                   protein molecule},
  volume =        {93},
  year =          {2004},
  doi =           {10.1103/PhysRevLett.93.180603},
}

@article{Schweizer1989,
  author =        {Schweizer, Kenneth S.},
  journal =       {J. Chem. Phys.},
  month =         {11},
  number =        {9},
  pages =         {5802-5821},
  title =         {Microscopic theory of the dynamics of polymeric
                   liquids: {General} formulation of a
                   mode–mode‐coupling approach},
  volume =        {91},
  year =          {1989},
  doi =           {10.1063/1.457533},
  issn =          {0021-9606},
}

@article{Schweizer1989a,
  author =        {Schweizer, Kenneth S.},
  journal =       {J. Chem. Phys.},
  number =        {9},
  pages =         {5822-5839},
  title =         {Mode‐coupling theory of the dynamics of polymer
                   liquids: {Qualitative} predictions for flexible chain
                   and ring melts},
  volume =        {91},
  year =          {1989},
  doi =           {10.1063/1.457534},
}

@article{Schweizer1991,
  author =        {Kenneth S. Schweizer},
  journal =       {J. Non-Cryst. Solids},
  pages =         {643-649},
  title =         {Mode-mode-coupling theory of the dynamics of
                   polymeric liquids},
  volume =        {131-133},
  year =          {1991},
  doi =           {https://doi.org/10.1016/0022-3093(91)90662-P},
  issn =          {0022-3093},
}

@article{Schweizer1997,
  author =        {Schweizer, Kenneth S. and Fuchs, Matthias and
                   Szamel, G. and Guenza, Marina and Tang, Hai},
  journal =       {Macromol. Theory Simul.},
  number =        {6},
  pages =         {1037-1117},
  title =         {Polymer-mode-coupling theory of the slow dynamics of
                   entangled macromolecular fluids},
  volume =        {6},
  year =          {1997},
  doi =           {https://doi.org/10.1002/mats.1997.040060604},
}

@article{Sharma2010,
  author =        {Sharma, Rati and Cherayil, Binny J.},
  journal =       {Phys. Rev. E},
  pages =         {021804},
  publisher =     {American Physical Society},
  title =         {Polymer melt dynamics: {Microscopic} roots of
                   fractional viscoelasticity},
  volume =        {81},
  year =          {2010},
  doi =           {10.1103/PhysRevE.81.021804},
}

@phdthesis{sharma2018theoretical,
  author =        {Sharma, Rati},
  school =   {Indian Institute of Science, Bangalore, India},
  title =         {Theoretical Approaches to the Study of Fluctuation
                   Phenomena in Various Polymeric Systems},
  year =          {2018},
  url =           {https://etd.iisc.ac.in/handle/2005/3338},
}

@article{Kargin1949,
  author =        {Kargin, V. A. and Slonimskii, G. L.},
  journal =       {Zh. Fiz. Khim.},
  number =        {5},
  pages =         {563--571},
  title =         {Determination of the molecular weight of linear
                   polymers from their mechanical properties
                   [Opredelenie molekulyarnogo vesa lineinykh polimerov
                   po ikh mekhanicheskim svoistvam]},
  volume =        {23},
  year =          {1949},
}

@phdthesis{Gotlib1952,
  address =       {Leningrad, USSR},
  author =        {Gotlib, Yu. Ya.},
  school =        {Leningrad State University},
  title =         {Thesis},
  year =          {1952},
}

@incollection{Likhtman2012,
  address =       {Amsterdam},
  author =        {A.E. Likhtman},
  booktitle =     {Polymer Science: A Comprehensive Reference},
  editor =        {Krzysztof Matyjaszewski and Martin Möller},
  pages =         {133-179},
  publisher =     {Elsevier},
  title =         {Viscoelasticity and Molecular Rheology},
  year =          {2012},
  doi =           {https://doi.org/10.1016/B978-0-444-53349-4.00008-X},
  isbn =          {978-0-08-087862-1},
}

@book{Ferry1980,
  address =       {{H}oboken, NY},
  author =        {Ferry, J.D.},
  edition =       {$3^\text{rd}$},
  publisher =     {John Wiley \& Sons Inc},
  title =         {{Viscoelastic properties of polymers}},
  year =          {1980},
}

@article{Caputo1967,
  author =        {Caputo, Michele},
  journal =       {Geophys. J. Int.},
  month =         {11},
  number =        {5},
  pages =         {529-539},
  title =         {Linear Models of Dissipation whose $Q$ is almost
                   Frequency Independent-{II}},
  volume =        {13},
  year =          {1967},
  doi =           {10.1111/j.1365-246X.1967.tb02303.x},
  issn =          {0956-540X},
}

@article{Everaers2020,
  author =        {Everaers, Ralf and Karimi-Varzaneh, Hossein Ali and
                   Fleck, Frank and Hojdis, Nils and Svaneborg, Carsten},
  journal =       {Macromolecules},
  number =        {6},
  pages =         {1901-1916},
  title =         {{Kremer-Grest} Models for Commodity Polymer Melts:
                   {Linking} Theory, Experiment, and Simulation at the
                   {Kuhn} Scale},
  volume =        {53},
  year =          {2020},
  doi =           {10.1021/acs.macromol.9b02428},
}

@article{Friedrich1991,
  author =        {Friedrich, C.},
  journal =       {Rheol. Acta},
  number =        {2},
  pages =         {151-158},
  title =         {Relaxation and retardation functions of the {Maxwell}
                   model with fractional derivatives},
  volume =        {30},
  year =          {1991},
  doi =           {10.1007/BF01134604},
  issn =          {1435-1528},
}

@book{RubinsteinPP,
  address =       {New York, NY},
  author =        {Rubinstein, Michael and Colby, Ralph H.},
  publisher =     {Oxford University Press},
  title =         {Polymer Physics},
  year =          {2003},
  isbn =          {978-0198520597},
}

@article{Santangelo2001,
  author =        {Santangelo, P. G. and Roland, C. M.},
  journal =       {J. Rheol.},
  month =         {03},
  number =        {2},
  pages =         {583-594},
  title =         {Interrupted shear flow of unentangled polystyrene
                   melts},
  volume =        {45},
  year =          {2001},
  doi =           {10.1122/1.1349711},
  issn =          {0148-6055},
}

@article{Poh2020,
  author =        {Poh, Leslie and Narimissa, Esmaeil and
                   Wagner, Manfred H.},
  journal =       {Rheol. Acta},
  month =         {Oct},
  number =        {10},
  pages =         {755-763},
  title =         {Universality of steady shear flow of Rouse melts},
  volume =        {59},
  year =          {2020},
  doi =           {10.1007/s00397-020-01236-2},
  issn =          {1435-1528},
}

@article{Oeser1988,
  author =        {Oeser, R. and Ewen, B. and Richter, D. and
                   Farago, B.},
  journal =       {Phys. Rev. Lett.},
  month =         mar,
  pages =         {1041--1044},
  publisher =     {American Physical Society},
  title =         {Dynamic fluctuations of crosslinks in a rubber: a
                   neutron-spin-echo study},
  volume =        {60},
  year =          {1988},
  doi =           {10.1103/PhysRevLett.60.1041},
}

@Article{Richter1989,
  author   = {D. Richter and B. Farago and B. Ewen and R. Oeser},
  title    = {The fluctuations of cross-links in a rubber — a neutron spin echo study—},
  journal  = {Physica B},
  year     = {1989},
  volume   = {156-157},
  pages    = {426--429},
  issn     = {0921-4526},
  doi      = {10.1016/0921-4526(89)90696-0},
  url      = {https://www.sciencedirect.com/science/article/pii/
                  0921452689906960},
}

@article{Gold2017,
  author =        {Gold, B. J. and H\"{o}velmann, C. H. and
                   L\"{u}hmann, N. and Székely, N. K. and
                   Pyckhout-Hintzen, W. and Wischnewski, A. and
                   Richter, D.},
  journal =       {ACS Macro Lett.},
  number =        {2},
  pages =         {73--77},
  title =         {Importance of compact random walks for the rheology
                   of transient networks},
  volume =        {6},
  year =          {2017},
  doi =           {10.1021/acsmacrolett.6b00880},
}

@article{Stockmayer1967,
  author =        {W. H. Stockmayer},
  journal =       {Pure Appl. Chem.},
  number =        {3-4},
  pages =         {539--554},
  title =         {Dielectric dispersion in solutions of flexible
                   polymers},
  volume =        {15},
  year =          {1967},
  doi =           {doi:10.1351/pac196715030539},
  url =           {https://doi.org/10.1351/pac196715030539},
}

@article{Watanabe2001,
  author =        {Watanabe, Hiroshi},
  journal =       {Macromol. Rapid Commun.},
  number =        {3},
  pages =         {127-175},
  title =         {Dielectric Relaxation of Type-{A} Polymers in Melts
                   and Solutions},
  volume =        {22},
  year =          {2001},
  doi =
  {https://doi.org/10.1002/1521-3927(200102)22:3<127::AID-MARC127>3.0.CO;2-S},
}

@article{Watanabe2002,
  author =        {Watanabe, Hiroshi and Matsumiya, Yumi and
                   Inoue, Tadashi},
  journal =       {Macromolecules},
  number =        {6},
  pages =         {2339-2357},
  title =         {Dielectric and Viscoelastic Relaxation of Highly
                   Entangled Star Polyisoprene:  {Quantitative} Test
                   of Tube Dilation Model},
  volume =        {35},
  year =          {2002},
  doi =           {10.1021/ma011782z},
}

@article{Tress2019,
  author =        {Tress, Martin and Xing, Kunyue and Ge, Sirui and
                   Cao, Pengfei and Saito, Tomonori and Sokolov, Alexei},
  journal =       {Eur. Phys. J. E},
  month =         oct,
  number =        {10},
  pages =         {133},
  title =         {What dielectric spectroscopy can tell us about
                   supramolecular networks},
  volume =        {42},
  year =          {2019},
  doi =           {10.1140/epje/i2019-11897-4},
  issn =          {1292-895X},
}

@article{Ge2020,
  author =        {Ge, Sirui and Tress, Martin and Xing, Kunyue and
                   Cao, Peng-Fei and Saito, Tomonori and
                   Sokolov, Alexei P.},
  journal =       {Soft Matter},
  pages =         {390--401},
  publisher =     {The Royal Society of Chemistry},
  title =         {Viscoelasticity in associating oligomers and
                   polymers: experimental test of the bond lifetime
                   renormalization model},
  volume =        {16},
  year =          {2020},
  doi =           {10.1039/C9SM01930H},
}

@article{Ge2023,
  author =        {Ge, Sirui and Carden, Gregory Peyton and
                   Samanta, Subarna and Li, Bingrui and Popov, Ivan and
                   Cao, Peng-Fei and Sokolov, Alexei P.},
  journal =       {Macromolecules},
  pages =         {to appear},
  title =         {Associating polymers in the strong interaction
                   regime: validation of the bond lifetime
                   renormalization model},
  year =          {2023},
  doi =           {10.1021/acs.macromol.2c02446},
}

@article{Arbe2023,
  author =        {Arbe, Arantxa and Alegría, Angel and Colmenero, Juan and
                   Bhaumik, Saibal and Ntetsikas, Konstantinos and
                   Hadjichristidis, Nikos},
  journal =       {ACS Macro Lett.},
  note =          {PMID: 37947419},
  number =        {11},
  pages =         {1595-1601},
  title =         {Microscopic Evidence for the Topological Transition
                   in Model Vitrimers},
  volume =        {12},
  year =          {2023},
  doi =           {10.1021/acsmacrolett.3c00586},
}

@article{Alegria2024,
  author =        {Alegría, Angel and Arbe, Arantxa and Colmenero, Juan and
                   Bhaumik, Saibal and Ntetsikas, Konstantinos and
                   Hadjichristidis, Nikos},
  journal =       {Macromolecules},
  number =        {12},
  pages =         {5639-5647},
  title =         {Segmental and Chain Dynamics of Polyisoprene-Based
                   Model Vitrimers},
  volume =        {57},
  year =          {2024},
  doi =           {10.1021/acs.macromol.3c02558},
}

@article{Watanabe2000,
  author =        {Watanabe, Hiroshi and Matsumiya, Yumi and
                   Osaki, Kunihiro},
  journal =       {J. Polym. Sci., Part B: Polym. Phys.},
  number =        {8},
  pages =         {1024-1036},
  title =         {Tube dilation process in star-branched
                   cis-polyisoprenes},
  volume =        {38},
  year =          {2000},
  doi =
  {https://doi.org/10.1002/(SICI)1099-0488(20000415)38:8<1024::AID-POLB3>3.0.CO;2-\#},
}

@article{Shanbhag2001,
  author =        {Shanbhag, S. and Larson, R. G. and Takimoto, J. and
                   Doi, M.},
  journal =       {Phys. Rev. Lett.},
  pages =         {195502},
  title =         {Deviations from Dynamic Dilution in the Terminal
                   Relaxation of Star Polymers},
  volume =        {87},
  year =          {2001},
}

@book{kremer2002broadband,
  address =       {Berlin, Heidelberg},
  author =        {Kremer, Friedrich and Sch{\"o}nhals, Andreas},
  edition =       {1st},
  publisher =     {Springer},
  title =         {{Broadband Dielectric Spectroscopy}},
  year =          {2002},
}

@article{Shanbhag2019respect,
  author =        {Shanbhag, Sachin},
  journal =       {Macromol. Theory Simul.},
  pages =         {1900005},
  title =         {{pyReSpect: A} computer program to extract discrete
                   and continuous spectra from stress relaxation
                   experiments},
  year =          {2019},
  doi =           {10.1002/mats.201900005},
}

@article{Virtanen2020,
  author =        {Virtanen, Pauli and Gommers, Ralf and
                   Oliphant, Travis E. and Haberland, Matt and
                   Reddy, Tyler and Cournapeau, David and
                   Burovski, Evgeni and Peterson, Pearu and
                   Weckesser, Warren and Bright, Jonathan and
                   {van der Walt}, St{\'e}fan J. and Brett, Matthew and
                   Wilson, Joshua and Millman, K. Jarrod and
                   Mayorov, Nikolay and Nelson, Andrew R. J. and
                   Jones, Eric and Kern, Robert and Larson, Eric and
                   Carey, C J and Polat, {\.I}lhan and Feng, Yu and
                   Moore, Eric W. and {VanderPlas}, Jake and
                   Laxalde, Denis and Perktold, Josef and
                   Cimrman, Robert and Henriksen, Ian and
                   Quintero, E. A. and Harris, Charles R. and
                   Archibald, Anne M. and Ribeiro, Ant{\^o}nio H. and
                   Pedregosa, Fabian and {van Mulbregt}, Paul and
                   {SciPy 1.0 Contributors}},
  journal =       {Nat. Methods},
  pages =         {261--272},
  title =         {{{SciPy} 1.0: Fundamental Algorithms for Scientific
                   Computing in Python}},
  volume =        {17},
  year =          {2020},
  doi =           {10.1038/s41592-019-0686-2},
}

@article{Shanbhag2023_psi,
  author =        {Shanbhag, Sachin},
  journal =       {J. Rheol.},
  month =         {07},
  number =        {5},
  pages =         {965-975},
  title =         {{A computer program for interconversion between creep
                   compliance and stress relaxation}},
  volume =        {67},
  year =          {2023},
  doi =           {10.1122/8.0000695},
  issn =          {0148-6055},
}

@Article{Shanbhag2026,
  author   = {Shanbhag, Sachin},
  title    = {Creep ringing in generalized linear viscoelastic models},
  journal  = {Rheol. Acta},
  year     = {2026},
  volume   = {65},
  pages    = {677-692},
  doi      = {10.1007/s00397-026-01577-4},
}

@article{Perego2022,
  author =        {Perego, Alessandro and Lazarenko, Daria and
                   Cloitre, Michel and Khabaz, Fardin},
  journal =       {Macromolecules},
  number =        {17},
  pages =         {7605--7613},
  title =         {Microscopic dynamics and viscoelasticity of
                   vitrimers},
  volume =        {55},
  year =          {2022},
  doi =           {10.1021/acs.macromol.2c00588},
}

@article{Song2023,
  author =        {Song, Jake and Holten-Andersen, Niels and
                   McKinley, Gareth H.},
  journal =       {Soft Matter},
  month =         {11},
  number =        {41},
  pages =         {7885-7906},
  title =         {Non-{Maxwellian} viscoelastic stress relaxations in
                   soft matter},
  volume =        {19},
  year =          {2023},
  doi =           {10.1039/d3sm00736g},
  issn =          {1744-683X},
}

\end{document}